\documentclass[12pt]{JHEP3}

\usepackage{epsfig}
\epsfclipon
\usepackage{multicol}
\usepackage{epsfig,bm}
\usepackage{amssymb,amsmath}

\newcommand{\roughly}[1]{\mathrel{\raise.3ex\hbox{$#1$\kern-0.85em
\lower1ex\hbox{$\sim$}}}}

\def\be{\begin{equation}}
\def\beq\begin{equation}
\def\ee{\end{equation}}
\def\bea{\begin{eqnarray}}
\def\eea{\end{eqnarray}}

\def\pref#1{(\ref{#1})}

\def\beq{\begin{equation}}
\def\eeq{\end{equation}}
\def\beqa{\begin{eqnarray}}
\def\eeqa{\end{eqnarray}}

\def\cF{{\cal F}}

\newcommand{\bmat}{\left(\begin{array}}
\newcommand{\emat}{\end{array}\right)}

\def\yzero{\smash{\hbox{$y\kern-4pt\raise1pt\hbox{${}^\circ$}$}}}

\def\-{\hphantom{-}}

\def\s2{\frac{1}{2}}

\def\tr{{\rm tr \,}}
\def\Tr{{\rm Tr \,}}

\def\IF{\relax{\rm I\kern-.18em F}}
\def\II{\relax{\rm I\kern-.18em I}}
\def\IP{\relax{\rm I\kern-.18em P}}
\def\IC{\relax{\rm I\kern-.48em C}}
\def\IR{\relax{\rm I\kern-.18em R}}
\def\IK{\relax{\rm I\kern-.20em K}}
\def\IM{\relax{\rm I\kern-.25em M}}

\def\cA{{\cal A}}

\def\cN{{\cal N}}

\def\Dsl{\,\raise.15ex\hbox{/}\mkern-13.5mu D} 
\def \one{\relax{\rm 1\kern-.26em I}}

\def\exd{{\rm d}}

\def\nott#1{\setbox0=\hbox{$#1$}                
   \dimen0=\wd0                                 
   \setbox1=\hbox{/} \dimen1=\wd1               
   \ifdim\dimen0>\dimen1                        
      \rlap{\hbox to \dimen0{\hfil/\hfil}}      
      #1                                        
   \else                                        
      \rlap{\hbox to \dimen1{\hfil$#1$\hfil}}   
      /                                         
   \fi}                                         %

\def\L{\mathcal{L}}
\def\g{\sqrt{-g}}
\def\Y{Y_{MN}}
\def\y2{Y_{MN} Y^{MN}}
\def\Riem2{R_{ABMN} R^{ABMN}}
\def\Ricci2{R_{MN} R^{MN}}

\def\f2{F^{a}_{MN} F^{MN}_a}
\def\nn{\nonumber}

\def\Y{Y^m_l}
\def\Ycc{Y^{m*}_l}
\def\Ypr{Y^{m'}_{l'}}

\newcommand{\oneb}{\hbox{\bf 1}}
\newcommand{\twob}{\hbox{\bf 2}}
\newcommand{\threeb}{\hbox{\bf 3}}
\newcommand{\fourb}{\hbox{\bf 4}}
\newcommand{\fiveb}{\hbox{\bf 5}}

\newcommand{\eightb}{\hbox{\bf 8}}

\newcommand{\tenb}{\hbox{\bf 10}}
\newcommand{\twelveb}{\hbox{\bf 12}}
\newcommand{\fourteenb}{\hbox{\bf 14}}
\newcommand{\sixteenb}{\hbox{\bf 16}}
\newcommand{\twentyfourb}{\hbox{\bf 24}}
\newcommand{\sixtyfourb}{\hbox{\bf 64}}
\newcommand{\eightyb}{\hbox{\bf 80}}

\newcommand{\ra}{\rightarrow}

\newcommand{\cR}{{\cal R}}

\newcounter{oldcounter}
\addtocounter{equation}{1}
\title{UV Sensitivity in Supersymmetric Large Extra Dimensions:
    The Ricci-flat Case}

\author{C.P. Burgess${}^{1,2,3}$ and D. Hoover${}^1$\\

${}^1$ Physics Department, McGill University,\\
\qquad 3600 University Street,
 Montr{\'e}al, Qu{\'e}bec, Canada, H3A 2T8. \\
${}^2$ Department of Physics and Astronomy, McMaster University,\\
\qquad 1280 Main Street West, Hamilton, Ontario, Canada, L8S 4M1.\\
${}^3$ Perimeter Institute,\\ \qquad 31 Caroline Street North,
Waterloo, Ontario, Canada. }

\date{}

\abstract{We examine the ultraviolet (UV)-sensitive part of the
one-loop Casimir energy which is induced when various
higher-dimensional supergravities are compactified to 4D on extra
dimensions which are Ricci flat, but otherwise arbitrary. We
identify the leading dependence on the mass of very massive
higher-dimensional modes, as well as the UV divergent part of the
contributions of modes which are massless in the
higher-dimensional sense (but which consist of a KK tower of
massive modes from the 4D perspective), and show how these are
constrained by higher-dimensional general covariance. Some of the
implications of co-dimension 2 branes are computed in the limit
where their tension is small compared with the extra-dimensional
Planck scale (but not small compared with the observed dark
energy). Our results support the interpretation of supersymmetric
large extra dimensions (SLED) in 6 dimensions as a potential
solution to the cosmological constant problem (but do not yet
completely clinch the case).\\

{\it This article is dedicated to Bryce DeWitt --- a pioneer in
the heat-kernel techniques used here --- whose recent passing
deprives physics of a rare mind.}}

\preprint{McGill-04/24.}


\begin{document}



\section{Introduction and Conclusions}

The long-standing problem of understanding why the energy density
of the vacuum is unobservably small \cite{ccreview} has recently
been recast as the problem of why it is extremely small but has
the nonzero value $\rho_{\rm vac} \approx (3 \times 10^{-3}$
eV$)^4$ \cite{DEdiscovery}. The most challenging part of this
problem is to understand why such a small energy density should be
`technically' natural, in the precise sense most clearly
enunciated in another context in ref.~\cite{technicalnaturalness}.
It should be emphasized that throughout the history of science
{\it all} of the many hierarchies of scale which have been
encountered --- at energies to which we have experimental access
--- are understood in a technically natural way {\it except} the
vacuum energy density.

In essence, a small parameter (like the vacuum energy density) is
technically natural if its smallness can be understood within the
effective theory which describes physics at any scale at which one
cares to pose the question. That is, even given that the vacuum
energy were small in some microscopic theory of the physics of
very short distances, why should it {\it remain} small as the
physics describing longer distances is integrated out? In
particular, since integrating out a particle of mass $m$ generates
a contribution to $\rho_{\rm vac}$ which is of order
$m^4$,\footnote{A positive power of the mass rather than the
cutoff appears here due to our use throughout of dimensional
regularization. The well-known issues of how to understand
power-law divergences in dimensional regularization is discussed
in Appendix A.} what cancels the huge contributions which are
obtained when all of the known particles (whose masses range
between $m_e \sim 0.5$ MeV and $m_t \sim 180$ GeV) are integrated
out? Failure to understand this point seems to indicate that our
formulation of physics misses something important at energies
above the vacuum-energy scale, $v \sim 10^{-3}$ eV, but in a way
which somehow affects only gravitational phenomena.

It may be that progress is being made on this thorny problem,
based on the recent recognition that reasonable theories of short
distance physics (like string theory) can predict that all of the
observed non-gravitational particles are trapped on
domain-wall-like surfaces, or branes, which are embedded within
spacetime \cite{DBranes}. Such a picture has already dramatically
broadened our perspective as to how the low-energy world could
look, such as by introducing the possibility that extra dimensions
could be as large as $r \sim 100$ microns across and yet have
hitherto escaped detection due to our only being able to search
for their existence using gravitational probes. It turns out that
such a picture can be viable, although consistency with the
observed value of Newton's constant only permits precisely 2
dimensions to be this large \cite{ADD}. Intriguingly, since $r
\sim 100$ $\mu$m implies $\hbar c/r \sim 10^{-3}$ eV, having all
presently-observed particles trapped on a 3-brane living within a
6-dimensional spacetime has the desired effect of dramatically
changing gravitational physics at energies above $10^{-3}$ eV,
without appreciably changing other non-gravitational phenomena at
observable scales.

This observation underlies a recent proposal for understanding the
smallness of the vacuum energy density within the framework of
6-dimensional supergravity with the two internal dimensions being
sub-millimeter in size: Supersymmetric Large Extra Dimensions
(SLED) \cite{Towards,SLEDrefs,SLEDreviews,SLEDrelated,CV2}. In
this picture the large extra dimensions contribute in two ways to
the explanation of the smallness of the observed 4D vacuum energy.
First, although quantum loops involving the observed particles do
give large contributions --- $O(M^4)$ for $M \sim \hbox{TeV}$
--- to the 4D vacuum energy, this energy is localized at the
position of the branes on which these particles reside. In many
situations the classical gravitational response of the internal
dimensions to this localized energy density appears to
systematically cancel the brane tensions in the effective 4D
vacuum energy to which cosmology is sensitive.\footnote{It should
be emphasized that the completeness of this cancellation is still
under active study. See, for example, ref.~\cite{CV2} for a recent
germane analysis.} In this paper we have nothing to say about the
on-going investigation of this classical response of the bulk.

The second way that the extra dimensions contribute to the
effective 4D vacuum energy density within the SLED proposal is
through their quantum response to --- or Casimir energy due to ---
the presence of the branes on which ordinary particles live. Given
the above-mentioned cancellation of the brane energies with the
classical bulk response, it is this quantum response of the bulk
which would make up the entirety of the observed Dark Energy
density. In particular, it has been argued
\cite{Towards,SLEDreviews} that within supersymmetric theories
this Casimir energy can be naturally as small as $1/r^4$, and so
account for the cosmological observations provided $r$ is of order
100 $\mu$m.\footnote{Once factors of $2\pi$ are included it turns
out that $r \sim 10$ $\mu$m, and so $\hbar c/r \sim 0.01$ eV, is
required for agreement with the observed Dark Energy density
\cite{SLEDrefs}.}

In this paper we address ourselves to testing whether the quantum
response of the extra dimensions can really be this small even
though there are large, $O(M)$, bulk states which circulate within
loops. In particular, although various one-loop Casimir energies
for 6D theories compactified to 4D exist \cite{Casimir6D} and the
results {\it are} of order $1/r^4$, these calculations typically
only integrate out massless fields in the bulk. However we know
that in 4 dimensions it is the integration over the most massive
fields which contribute the largest amount to the vacuum energy
(recall that contributions to the bulk cosmological constant vary
like $m^n$ in $n$ dimensions), so we focus here on the quantum
effects of bulk fields whose mass satisfies $M \gg 1/r$. In
general the Casimir energy (per unit 4-volume) of such fields can
be expected to depend on both $M$ and $r$, and for $M \gg 1/r$
could generically have the form:
\begin{equation}
    V_C(M,r) = c_0 M^6 r^2 + c_1 M^4 +  \frac{c_2 M^2}{r^2}
    + \frac{c_3}{r^4} + \dots \,.
\end{equation}
Here the leading contribution depends on $M$ and $r$ as does the
contribution of a 6D cosmological constant. The dimensionless
constants, $c_k$, are calculable for a given field content in the
bulk and on the branes, and the success of the SLED proposal
requires the conditions $c_0 = c_1 = c_2 = 0$. Our goal in this
paper is to track these positive powers of $M$ in the Casimir
energy at one loop, to see whether or not they really do vanish.

Fortunately, very general tools exist, based on heat-kernel
methods \cite{Gilkey,deWitt,Duff,HKReview}, for determining the
large-$M$-dependence of the Casimir energy, for a broad choice of
fields within the bulk and for a wide variety of geometries for
the extra dimensions. We here adapt these techniques to 6D field
theories containing scalars, fermions, gauge bosons, gauge
2-forms, gravitini and gravitons, which are compactified to 4D on
an arbitrary Ricci-flat manifold. The restriction to
Ricci-flatness is introduced for technical convenience (and is
relaxed in a companion paper \cite{Doug2}). Due to the absence of
a bulk cosmological constant in 6D supergravity theories this
restriction essentially precludes the study of compactifications
in the presence of fluxes, and with spatially-varying scalar
fields (which are indeed among the more interesting 6D
compactifications \cite{SLEDrefs,SLEDrelated,CV2} for the chiral,
gauged supergravities in 6 dimensions). Ricci-flatness remains of
interest, however, because it does include a broad class of
nontrivial solutions to many 6D supergravities.

In the heat-kernel treatment all of these dangerous large-$M$
terms arise as local effective interactions within the bulk 6D
theory. We use these techniques to identify these effective
interactions in a general way for the 6D fields of most interest.
In general bulk loops can also generate large-$M$ effective
interactions which are localized at the positions of the branes.
We compute the leading dependence of these terms in the limit when
the brane tensions are smaller than the 6D gravitational scale
(but are much larger than $1/r$).

Our final results are as follows. We find that the Casimir energy
due to any one 6D field typically does involve terms proportional
to $M^6$, $M^4$ or $M^2$. However our main result is to find that
all three of these kinds of terms cancel once they are summed over
a massive supermultiplet of $(2,0)$ supersymmetry in 6 dimensions.
Furthermore, this cancellation is independent of the details of
the compactification from 6D to 4D (provided it is Ricci flat),
and in particular does {\it not} require that the compactification
be supersymmetric. This result applies in particular to the
contributions of the massive Kaluza-Klein (KK) modes which would
be obtained if the 6D supergravity of interest were itself
obtained from a 10- (or 11-) dimensional theory by dimensional
reduction on 4 (or 5) small dimensions, since these have the same
particle content as do the massive $(2,0)$ 6D supermultiplets we
examine.

As such, we believe these results to be encouraging for the
success of the SLED program, inasmuch as they show how bulk
supersymmetry can stop very heavy KK particles from contributing
too large one-loop amounts to the low-energy effective
cosmological constant, both through bulk and brane-localized
effective interactions. These results leave one type of one-loop
contribution which could still be dangerously large from the SLED
point of view: dimension-two interactions localized on the brane
(such as a brane-localized Einstein-Hilbert action) which are
generated by loops of brane-bound particles (for which
supersymmetry is badly broken). Indeed, such effective
interactions are likely to be generated by brane loops, and it is
not yet clear \cite{CV2} whether the bulk response need cancel
their 4D effects in the same way as it would do for an effective
brane tension.

The rest of the paper is organized as follows. Section 2 collects
many of the general results on UV sensitivity and Kaluza-Klein
theories which prove relevant to the 6D discussion. Section 3 then
uses these results to compute explicitly the UV-sensitive
contributions of various 6D massless and massive particles, and
assembles these results into supermultiplets to see how the
presence of higher-dimensional supersymmetry changes the generic
picture.

\section{General Results}

Our goal is to compute explicitly (as functions of background
fields) the ultra-violet sensitive part of the one-loop vacuum
energy for compactifications to 4D of various 6D field theories.
Before embarking on the full calculation it is worth first
collecting a few general results concerning the kinds of
ultraviolet divergences which can be encountered in calculations
of this sort. Because the ultraviolet behavior only depends on the
very short-distance limit of the theory these divergences can
always be absorbed into renormalizations of local functions of the
background fields, with coefficients which can be computed very
generally for arbitrary background geometries
\cite{Gilkey,deWitt,Duff,HKReview}.

\subsection{The Gilkey-DeWitt Coefficients}

This section collects the results for the ultraviolet-divergent
parts of the one-loop action obtained by integrating out various
kinds of particles in 6 dimensions. To this end, consider a
collection of $N$ fields, $\Psi^z$ with $z = 1,...,N$, coupled to
a collection of background fields, possibly including a
6-dimensional spacetime metric, $g_{MN}$, scalars, $\varphi^i$,
and form fields, $A^a_{M_1..M_p}$. For each $z$, $\Psi^z$ can
carry a gauge and/or Lorentz index, although for simplicity of
notation the Lorentz index is suppressed in this section. We
suppose that the background-covariant derivative, $D_M$,
appropriate to $\Psi^z$ is:
\be
    D_M \Psi^z = \partial_M \Psi^z + \omega_M \, \Psi^z
    - i A^a_M \, {(t_a)^z}_y \Psi^y
     \,,
\ee
where $\omega_M$ is the appropriate matrix-valued spin connection,
and the gauge group is represented by the hermitian matrices
${(t_a)^z}_y$. For real fields the $t_a$ are imaginary
antisymmetric matrices, and (for canonically-normalized gauge
bosons) we take the gauge group generators to include a factor of
the corresponding gauge coupling, $g_a$. The commutator of two
such derivatives defines a generalized matrix-valued curvature,
${(Y_{MN})^z}_y \Psi^y = [D_M,D_N] \Psi^z$, which has the
following form:\footnote{We adopt Weinberg's curvature conventions
\cite{GandC}, which differ from those of MTW \cite{MTW} only in an
overall sign in the definition of the Riemann tensor.}
\be
    {(Y_{MN})^z}_y = \cR_{MN} \, {\delta^z}_y
    -i F^a_{MN} \, {(t_a)^z}_y \,.
\ee
Here $\cR_{MN}$ is the curvature built from the spin connection
$\omega_M$, which is also related to the Riemann curvature of the
background spacetime in a way which is made explicit in what
follows.

Quite generally the result of integrating out the fields $\Psi^z$
at one-loop leads to the following contribution to the effective
quantum action
\beq     \label{eqn: sigma}
    i\Sigma = - (-)^F \, \frac{1}{2} \, \Tr \log
    \Delta \,,
\eeq
where $(-)^F = 1$ for bosons and $-1$ for fermions, and the
differential operator ${\Delta^z}_y$ has the following form
\beq
    {\Delta^z}_{y} = - {\delta^z}_{y} \, \Box
    + {X^z}_{y} + {(m^2)^z}_{y}
    \,.
\eeq
Here ${\delta^z}_y$ is the Kronecker delta, $\Box = g^{MN} D_M
D_N$ and ${X^z}_{y}$ is a local quantity built from the background
fields whose form depends on the kind of field under consideration
(explicit examples are given below for the usual fields of
interest). The mass matrix, $m^2$, can either be regarded as being
physical masses which are extracted from within $X$, or as a
regulator mass, ${(m^2)^z}_{y} = \mu^2 \, {\delta^z}_{y}$, which
is to be taken to zero (or to infinity) at the end of the
calculation.

Our interest for this section is in two parts of $\Sigma$ which are
very closely related to one another. One of these is the ultraviolet
divergent part of $\Sigma$, and the other is that part of $\Sigma$
which depends most strongly on the mass of any massive 6D fields
which are integrated out. We collect here the very general results
which can be obtained for both of these quantities using the
Gilkey-DeWitt heat-kernel methods
\cite{Gilkey,deWitt,Duff,HKReview}. When identifying the divergent
part we work within dimensional regularization and so continue the
spacetime dimension to complex values, $n$, which are slightly
displaced from the actual integer spacetime dimension, $6$, which is
of interest: $n = 6 - 2\epsilon$. We then follow the poles in
$\Sigma$ as $\epsilon \to 0$, in the usual fashion. These may be
related to the logarithmic divergences which would be obtained from
an ultraviolet cutoff, $\Lambda$, through the usual relation (see
Appendix~\ref{DRvsCutoffs})
\be \label{divrelation}
    \frac{1}{\epsilon} \leftrightarrow \ln \left( \Lambda^2 \right) \,.
\ee

For 6D spaces without boundaries and singularities the
ultraviolet-divergent terms (and heavy-mass-dependent terms) are
simply characterized. In $n$ dimensions they may be written as
\cite{Gilkey,HKReview}
\be \label{sigmainftyndim}
    \Sigma_\infty = \frac{1}{2} (-)^F \left( \frac{1}{4\pi} \right)^{n/2}
    \int d^{\,n}x \sqrt{-g} \sum_{k=0}^{[n/2]} \Gamma(k-n/2)
    \Tr[m^{n-2k} \, a_k],
\ee
which for $n = 6 - 2\epsilon$ specializes to
\beq \label{sigmainfty}
    \Sigma_\infty =
    \frac{1}{2(4 \pi)^3} \, (-)^F
    \sum_{k=0}^{3} \Gamma ( k - 3 + \epsilon)
    \int d^6 x \sqrt{-g} \, \Tr[ m^{6-2k}\, a_k ] \,.
\eeq
Here $\Gamma(z)$ denotes Euler's gamma function. The divergence as
$\epsilon \to 0$ is contained within the gamma function, which has
poles at non-positive integers of the form $\Gamma(-r +\epsilon) =
(-)^r/(r!\epsilon)+\cdots$, for $\epsilon$ an infinitesimal and
$r$ a non-negative integer. The coefficients, $a_k$, are known $N'
\times N'$ matrix-valued local quantities constructed from the
background fields, to which we return below. Here $N' = N \, d$
with $N$ counting the number of fields and $d$ being the dimension
of the appropriate Lorentz representation. The trace is over the
$N'$ matrix indices of the $a_{k}$.

The above expression shows that for massless fields ($m = 0$) in
compact spaces without boundaries and singularities in 6
dimensions the divergent contribution is proportional to
$\tr[a_{3}]$ in dimensional regularization, so the problem of
identifying these divergences reduces to the construction of this
coefficient.\footnote{Notice that the freedom to keep $m^2$ within
or separate from $X$ implies that the divergence obtained from
computing just $a_3$ using $X_m = X + m^2$ gives the same result
as computing $a_0$ through $a_3$ using only $X$.} By contrast, for
massive fields there are divergences proportional to $\tr [ m^6
a_0]$, $\tr[ m^4 a_1]$, $\tr[ m^2 a_2]$ and $\tr[a_3]$, and these
are also the terms in $\Sigma$ which involve the highest powers of
$m$. For example, it turns out that $a_0$ is proportional to the
unit matrix, $I$, and so the term involving $a_0$ represents a
divergent and strongly $m$-dependent contribution to the 6D
cosmological constant, proportional to $\tr [m^6]$. Similarly,
since $a_1$ contains a term proportional to $R \, I$, where $R$ is
the background metric's Ricci scalar, $\tr[m^4 a_1]$ contains an
$m$-dependent renormalization of the Einstein-Hilbert action, and
so on.

What is attractive about the above results is that an algorithm
for constructing the coefficients $a_k$ is known for general $X$
and $D_M$, and the result for the first few has been computed
explicitly \cite{Gilkey,HKReview} and can be given as a closed form in
terms of $X$, background curvatures and the generalized curvature
${(Y_{MN})^z}_y$. The explicit results for the quantities $a_0$
through $a_3$ are given in their general form in
Appendix~\ref{GilkeyCoefficients}. These
allow the calculation of the most ultraviolet-sensitive
contributions from quantum loops for arbitrary theories in the
presence of very general background field configurations.

The remainder of this section specializes these results to the
various fields of interest for 6D supergravity theories. We take
the bosonic part of the action for these theories to be
\bea \label{scalaraction}
     S &=& - \int d^6x \sqrt{-g} \; \left[ \frac12 \, g^{MN} \,
     G_{ij}(\Phi) \, D_M \Phi^i D_N \Phi^j
     + V(\Phi) \right. \nonumber \\
     && \qquad\qquad\qquad\qquad \left. + \frac12 \, U(\Phi) \, R +
     \sum_p \frac{1}{2p!} \, W_p(\Phi) \, F^a_{M_1..M_p}
     F_a^{M_1..M_p}
     \right] \,,
\eea
where $\Phi^i$ denote the theory's scalar fields, and $F_{(p)} =
\exd \, A_{(p-1)} + \omega_p$ is a $p$-form field strength for a
$(p-1)$-form gauge potential, and $\omega_p$ is an appropriate
Chern-Simons form whose details are not important in what follows.
The coefficient functions $U$, $V$, $W_p$ and $G_{ij}$ are known
functions of the $\Phi^i$ which differ for different choices for
the 6D supergravity of interest.

As usual, we are always free to use the classical equations of
motion obtained from this action to simplify the one-loop quantity
$\Sigma$, because anything which vanishes with the classical field
equations may be removed from $\Sigma$ by performing an appropriate
field redefinition \cite{irrelevant,ETbooks}. For simplicity we also
specialize here to the case where these classical equations are the
same as those for pure gravity, $R_{MN} = 0$, and so for which all
other fields have trivial stress energy: $F_{M_1..M_p} =
\partial_M \Phi^i = 0$. We return to the more general case with
more complicated classical solutions in a companion paper
\cite{Doug2}.

\subsection{Dimensional Reduction from 6D to 4D}

When using the Gilkey coefficients in 6-dimensional theories
compactified to 4 dimensions one might be tempted to ask whether
we should take $n=4$ or $n=6$ when evaluating formulae like
eq.~(\ref{sigmainftyndim}). In this section we show that it makes
no difference, inasmuch as the sum over the result for each 4D KK
mode reproduces the full 6D expression.

\subsubsection*{Dimensional Reduction on $S^2$}

To establish this point we take the simplest nontrivial example:
the reduction of a 6D scalar field theory to 4D on a
2-sphere.\footnote{Compactification on a torus is too trivial for
the present purposes, since all of the Gilkey coefficients except
$a_0$ tend to vanish for flat manifolds like torii. Ref.
\cite{Casimir6D} includes more recent explicit examples.} For
these purposes we start with the 6D action
\be
    S = -\int d^{\, 4}x d^{\, 2}y \sqrt{-g}( -\Phi^* \Box_6 \Phi +
    M^2 \Phi^* \Phi),
\ee
where $\Phi$ is a minimally-coupled complex scalar field with a 6D
mass $M$, and $\Box_6 = g^{MN} D_M D_N$ is the 6D d'Alembertian.

We further assume that the background metric takes the product
form $\exd s^2_6 = g_{\mu \nu}(x) \, \exd x^\mu \exd x^\nu +
g_{mn}(y) \, \exd y^m \exd y^n$, where $g_{mn}$ is the internal
two dimensions and $g_{\mu\nu}$ is the metric of the `large' 4
dimensions. In this case the 6D d'Alembertian is related to its 4D
counterpart, $\Box_4 = g^{\mu\nu} D_\mu D_\nu$, and the 2D
Laplacian, $\Box_2 = g^{mn} D_m D_n$, by $\Box_6 = \Box_4 +
\Box_2$. Finally, we specialize to an internal $S^2$ by taking
$g_{mn} \exd y^m \exd y^n = r^2 \, \gamma_{mn} \exd y^m \exd y^n =
r^2 (\exd \theta^2 + \sin^2 \theta \exd \varphi^2)$, where $r$
denotes the 2-sphere's radius.

The dimensional reduction is performed by writing $\Phi$ as a mode
sum in terms of the eigenfunctions of the scalar Laplacian on a
2-sphere. Our ansatz therefore becomes:
\be
    \Phi(x,y) = \frac{1}{r}
    \sum_{l=0}^\infty \sum_{m = -l}^l \phi_l^m(x) \Y(y),
\ee
where $\Y(y)$ are the standard spherical harmonics, and
$\phi_l^m(x)$ are the corresponding 4D fields. Using $\Box_6 =
\Box_4 + \Box_2$ we find
\be \label{kineticterm}
    \Phi^* \Box_{6} \Phi = \frac{1}{r^2} \sum_{m,l,m',l'}
    \left[ \Bigl(\phi_l^{m*}(x) \Box_{4} \phi^{m'}_{l'}(x)
    \Bigr) \Ycc \Ypr +
    \phi_l^{m*}(x) \phi_{l'}^{m'}(x) \Ycc \Box_2 \Ypr \right].
\ee

Finally, using the results $-\Box_2 \Y = [l(l+1)/r^2] \Y$,
$\sqrt{-g_6} = r^2 \, \sqrt{-g_4} \, \sqrt{\gamma}$ as well as the
orthonormality relations $\int d^2y \sqrt{\gamma} \, \Ycc \Ypr =
\delta_{ll'} \delta_{mm'}$, the 6D action becomes
\be
    S = - \sum_{m,l} \int d^{\, 4}x \sqrt{-g_4} \left[ -\phi_l^{m*}
    \Box \phi^m_l + \left( M^2 + \frac{l(l+1)}{r^2} \right)
    \phi_l^{m*} \phi_l^m \right],
\ee
where all quantities are now functions only of $x$. This is the
standard manipulation which expresses the theory of one complex 6D
scalar in terms of an infinite tower of 4D Kaluza-Klein (KK)
modes, consisting of complex 4D scalars whose 4D masses are
$\mu^2_l = M^2 + l(l+1)/r^2$. Notice that for scalars on the
2-sphere each KK mass level has degeneracy $d_l = (2l+1)$.

\subsubsection*{UV Sensitivy}

We now check that the UV sensitive terms are identical when
computed in 6 dimensions or as the sum over a series of 4D results
for each KK mode. Recall for these purposes that the divergent
part of the one-loop quantum action can be written in $n$
dimensions as
\be
    \Sigma_\infty = \frac{1}{2} (-)^F \left( \frac{1}{4\pi} \right)^{n/2}
    \int d^{\,n}x \sqrt{-g} \sum_{k=0}^{[n/2 \,]} \Gamma(k-n/2)
    \Tr[M^{n-2k} \, a_k],
\ee
where $M$ is the $n$-dimensional mass of the particle which
traverses the loop.

\medskip\noindent{\it The 6D Calculation:}
For the 6D calculation we use the general result specialized to a
minimally-coupled scalar field in $n = 6$ dimensions. For
simplicity we also assume the 6D complex scalar to be massless
--- so $M=0$ --- and take $Y_{MN} = X = 0$. Because the scalar
has been taken to be massless in 6D, the only relevant Gilkey
coefficient is $a_3$, which we must evaluate. For this evaluation
we specialize the general result to the product geometry, for
which $R_6 = R_4 + R_2$, $R_{MN} R^{MN} = R_{\mu\nu} R^{\mu\nu} +
R_{mn} R^{mn}$ {\it etc.}, where the 2-sphere curvatures satisfy
$R_{mnpq} = (1/r^2) (g_{mq} g_{np} - g_{mp} g_{nq})$, $R_{mn} =
-(1/r^2) g_{mn}$ and $R_2 = -2/r^2$, so $R_{mnpq} R^{mnpq} = 2
R_{mn} R^{mn} = R_2^2 = 4/r^4$.

Remembering the overall factor of 2 because the scalar is complex,
and expanding $a_3$ in powers of the 4D curvature tensor we find
\be
    \Tr[a_3] = \frac{8}{315r^6} - \frac{1}{45 r^4} \, R_4 + \dots
    \,,
\ee
and so integrating over the 2-sphere, and using $\Gamma(0) \sim
1/\epsilon$, we find
\be
    \Sigma_\infty = \frac{1}{2} \left( \frac{1}{4\pi} \right)^2
    \frac{1}{\epsilon} \int d^{\,4}x \sqrt{-g} \left( \frac{8}{315r^4}
    - \frac{1}{45 r^2} \, R_4 + \dots \right)
    .
\ee

\medskip\noindent{\it The 4D Calculation:}
In the 4D theory we may similarly take $X=0$ provided we separate
explicitly the KK mass terms from $X$, and up to linear order in
$R_4$ we need keep only the contributions to $a_1$ and $a_0$.
Using $\Gamma(-r + \epsilon) \sim (-)^r/(r! \epsilon)$ we find for
each KK mode (remembering again the factor of 2 for complex
scalars),
\bea
    \Sigma^{lm}_\infty &=& \frac{1}{2} \left( \frac{1}{4\pi} \right)^2 \int
    d^{\,4}x \sqrt{-g} \,\left( \Tr[ \mu_l^4 a_0] \, \frac{1}{2\epsilon}
    - \Tr[ \mu^2_l a_1] \, \frac{1}{\epsilon} + \dots \right) \nonumber\\
    &=& \frac{1}{2} \left( \frac{1}{4\pi} \right)^2 \int
    d^{\,4}x \sqrt{-g} \,\left( \frac{\mu_l^4}{\epsilon}
    + \frac{\mu^2_l}{3 \epsilon} \, R_4  + \dots \right) .
\eea

We now sum over the KK modes, and interpret the resulting
divergent sums using $\zeta$-function regularization
\cite{hardy,zeta,BD}. Recalling that each mass eigenvalue $\mu_l^2
= l(l+1)/r^2$ has degeneracy $(2l+1)$ we have
\bea
    \sum_{lm} \mu_l^4 &=& \sum_{l=0}^{\infty} \left( \frac{l(l+1)}{r^2}
    \right)^2 (2l+1) \nn \\
    &=& \frac{1}{r^4} \sum_{l=1}^{\infty} (2l^5 + 5l^4 + 4l^3 + l^2)
    \nn \\
    &=& \frac{1}{r^4} \Bigl[ 2 \zeta(-5) + 5 \zeta(-4) + 4 \zeta(-3) +
    \zeta(-2) \Bigr],
\eea
where $\zeta(s) = \sum_{n=1}^{\infty} (1/n^s)$ is the Riemann
zeta-function. Using the following results \cite{gr}
\bea
    && \zeta(-5) = -\frac{1}{252}, \ \zeta(-3) = \frac{1}{120}, \\
    && \zeta(-4) = \zeta(-2) = 0, \ \zeta(-1) = - \frac{1}{12} \,,
\eea
we find that
\be
    \sum_{lm} \mu_l^4 = \frac{8}{315 r^4}.
\ee
Similarly
\bea
    \sum_{lm} \mu_l^2 &=& \sum_{l=0}^{\infty} \frac{l(l+1)}{r^2}
     (2l+1) \nn \\
    &=& \frac{1}{r^2} \sum_{l=1}^{\infty} (2l^3 + 3l^2 + l)
    \nn \\
    &=& \frac{1}{r^2} \Bigl[ 2 \zeta(-3) +
    3 \zeta(-2) + \zeta(-1) \Bigr], \nn\\
    &=& - \frac{1}{15 \, r^2} \,.
\eea

Finally, combining these results we obtain the following
expression for the divergent piece, as computed in 4 dimensions:
\bea
    \Sigma_\infty &=& \sum_{lm} \Sigma^{lm}_\infty \nn\\
    &=& \frac{1}{2} \left( \frac{1}{4\pi} \right)^2
    \frac{1}{\epsilon} \int d^{\,4}x \sqrt{-g}
    \left( \frac{8}{315r^4} - \frac{1}{45} \, R_4 +
    \dots \right).
\eea
As expected, we obtain the same result for $\Sigma_\infty$
regardless of whether we do the calculation in the 6D or the 4D
theory, provided we sum over all of the KK modes in the
lower-dimensional case. It is therefore a matter only of
convenience whether or not to use the higher- or lower-dimensional
formulation.

\subsubsection*{Dimensional Reduction in Supersymmetric Models}

The previous calculations are useful when computing
the UV sensitivity of 6D supersymmetric theories, particularly
when 6D supersymmetry breaks due to the compactification down to 4
dimensions. Seen from the 4D point of view it might appear that
supersymmetry is badly broken, making the cancellations due to 6D
supersymmetry hard to follow. However, the freedom to perform
computations in the higher dimensions makes it easier to see the
cancellations which follow from higher-dimensional supersymmetry.
Physically these cancellations still hold because it is the UV
sensitive part of the one-loop result which we compute, and this
is only sensitive to the very short wavelengths for which the
higher-dimensional symmetries apply.

We now turn to a discussion those UV-sensitive effects which are
localized near the position of any co-dimension 2 branes.

\subsection{Brane-Localized Terms}

For supersymmetric large extra dimensions we require the Casimir
energy in the presence of brane sources, which typically
introduces either boundaries or singularities into the bulk
geometry, depending on the dimension of the brane involved. Since
the presence of boundaries and singularities permit the appearance
of more complicated divergences in the Casimir energy, additional
local counter-terms are required in order to renormalize them.
Since all of these are localized at the brane positions, they can
be regarded as renormalizations of the effective brane actions.

Unfortunately, results with the generality described above are not
yet available in the presence of co-dimension 2 brane sources.
Some things are known, however, and we summarize those which are
most relevant to the SLED proposal here. The main calculations
which have been done assume the geometry near the branes to be
described by a conical singularity, for which the 2D bulk metric
can be written in the form $\exd r^2 + c_b^2 r^2 \exd \theta^2$ near
the singularity ($r = 0$), where $\theta$ is a periodic coordinate
with period $2 \pi$ and $c_b$ is a constant. This geometry has a
defect angle at the brane position, whose size is given by $\delta
= 2 \pi(1-c_b)$. This introduces a delta-function-type divergence
into the curvature at the brane position which is proportional to
$\delta$. This kind of singularity is often (but not always
\cite{cosmicstringsnonconical,GGPplus}) what is produced by
3-branes which are aligned within the 6 dimensions to be parallel
with the large 4 dimensions.

Some explicit results are known for the types of ultraviolet
divergences which arise in this case. This includes explicit
results for the heat kernel coefficients for specific types of
particles in the presence of these singularities
\cite{HKReview,conicaldivergences} as well as more general
expressions which apply to the limit of small defect angles, which
are obtained by interpreting the cone to be the limit of a
sequence of `blunted' cones for each of which the tip is smoothed
off \cite{smalldefect,smalldefecthispin}. According to this line
of argument, the leading contributions (for small defect angles)
to the brane counter-terms may be found by taking the limit of the
bulk terms obtained for each of the blunted cones.

Applied to quadratic order in the background curvatures, this
leads to the relations:
\bea \label{Btoblimits}
    \int_{\tilde{B}} d^6x \, \sqrt{-g} \, \tilde{R} &\approx& \int_{B'} d^6x
    \, \sqrt{-g} \, R - \sum_b 4 \pi ( 1 - c_b) \int_b d^4x \,
    \sqrt{-h} \nonumber \\
    \int_{\tilde{B}} d^6x \, \sqrt{-g} \, \tilde{R}^2 &\approx& \int_{B'} d^6x
    \, \sqrt{-g} \, R^2 - \sum_b 8 \pi \left( 1 - c_b \right)
    \int_b d^4x \, \sqrt{-h} \, R \nonumber \\
    \int_{\tilde{B}} d^6x \, \sqrt{-g} \, \tilde{R}_{MN}
    \tilde{R}^{MN} &\approx& \int_{B'} d^6x
    \, \sqrt{-g} \, R_{MN} \, R^{MN} \nonumber \\
    &&\qquad\qquad - \sum_b 4 \pi \left( 1 - c_b \right)
    \int_b d^4x \, \sqrt{-h} \, R_{aa}  \\
    \int_{\tilde{B}} d^6x \, \sqrt{-g} \,
    \tilde{R}_{MNLP} \tilde{R}^{MNLP} &\approx& \int_{B'} d^6x
    \, \sqrt{-g} \, R_{MNLP} R^{MNLP} \nonumber \\
    && \qquad\qquad - \sum_b 8 \pi \left( 1 - c_b \right)
    \int_b d^4x \, \sqrt{-h} \, R_{abab} \nonumber
    \,,
\eea
where the approximate equality indicates that terms of order $(1 -
c_b)^2$ are neglected on the right-hand side. In these expressions
$\tilde{B}$ denotes the limit of the sequence of blunted cones
(having curvature $\tilde{R}_{MNLP}$) which approach the bulk,
$B$, including the conical singularity, and $b$ denotes the 4D
world-surface of the brane, defined by the position of the conical
singularity. $B'$ denotes the bulk with the positions of the
conical singularities removed: $B' = B - \sum_b b$, and $R_{MNLP}$
is the curvature of this bulk in the limit of no singularity.
$h_{\mu\nu}$ denotes the induced metric on $b$, which we also
suppose to have no extrinsic curvature, and
\be \label{Rnormals}
    R_{aa} = \sum_{a=1}^2 R_{MN} \, n^M_a \, n^N_a ,
    \quad\hbox{and}\quad
    R_{abab} = \sum_{a,b = 1}^2 R_{MNPQ} \, n^M_a \, n^M_b \,
    n^P_a \, n^Q_b \,
\ee
where $n_i^M$ denote two mutually-orthogonal unit normals to the
appropriate brane world surface.

Because these expressions for the brane-localized contributions to
the heat kernel are obtained as limits of a sequence of bulk
contributions, they permit an easy generalization of the
expressions given in Appendix B to include brane-localized terms
in the limit of small defect angles. We now summarize the results
which are obtained in this way, which give the following
brane-localized counterterms:
\bea
  \int_{\tilde{B}} d^6x \sqrt{-g} \, \Tr[m^4 a_1]
  &\approx& \int_{B'} d^6x \sqrt{-g} \, \Tr[m^4 a_1]
  +\sum_b \frac{2\pi}{3} ( 1 - c_b) \int_b d^4x \sqrt{-h}
  \, \Tr[m^4] \nonumber \\
  \int_{\tilde{B}} d^6x \sqrt{-g} \, \Tr[m^2 a_2]
  &\approx& \int_{B'} d^6x \sqrt{-g} \, \Tr[ m^2 a_2] \\
  &&\qquad - \sum_b \frac{2 \pi}{3} (1 - c_b) \int_b d^4x \sqrt{-h} \,
  \left\{\Tr[m^2 \hat{X}] \phantom{\frac12} \right. \nn \\
  &&\qquad\qquad\qquad \left. +\frac{1}{30} \left(2 R_{abab} -  R_{aa} +
  5 R \right) \Tr[m^2]  \right\} \,.\nn
\eea
Here $\Tr[m^2 \hat{X}]$ is defined as follows: if $\Tr[m^2 X] = aR
+ b$ then $\Tr[m^2 \hat{X}] = 2aR + b$.

These expressions assume that the defect angles are small and that
the only singular bulk fields near the brane positions are the
curvatures. This latter assumption is a natural consequence of our
assumption of the vanishing of background quantities like $X$ and
$F^a_{MN} F_a^{MN}$, since this guarantees that these quantities
remain smooth there. They predict (for small defect angles) the
new brane-localized ultraviolet divergences which arise once
branes are inserted into the bulk space.

Where the explicit results can be compared with this small-defect
limit they agree.\footnote{It is claimed in
ref.~\cite{smalldefecthispin} that the small-defect result does
not agree with explicit calculations for the divergences produced
by integrating out spin-3/2 and spin-2 particles. However we
regard these conclusions to be suspect inasmuch as the obstruction
they find explicitly involves the contributions of pure-gauge
modes --- {\it i.e.} conformal Killing vectors and spinors --- and
this reference does not treat properly the contributions of the
ghosts which would be expected to cancel such modes.}

Finally, since our present interest is in whether the Gilkey
coefficients vanish when summed over the elements of a 6D
supermultiplet, it is useful to notice here that --- to linear
order in $(1-c_b)$ --- the vanishing of the UV sensitive terms on
the boundary is an automatic consequence of the vanishing of all
of the corresponding bulk terms from which they arise. It would
clearly be very useful to have more general calculations of these
quantities.

\section{Explicit Calculations of Bulk Loops}

This section specializes these general results for the
ultraviolet-divergent parts of the one-loop action to several
specific particle types which arise in 6 dimensional
supergravities. For technical reasons we specialize in what
follows to the case where all background gauge fields vanish and
--- with applications to higher-dimensional supergravity in mind
--- where the higher-dimensional cosmological constant vanishes.
In this situation the classical field equations imply the
background metric must satisfy $R_{MN} = 0$, and so we also
restrict to Ricci-flat metrics.

\subsection{6D Massless Particles}

We start by computing the results for 6D massless particles.
Although we specialize in the end to vanishing gauge fields and to
Ricci-flat metrics, for later convenience we quote our
intermediate results for lower-spin fields in the more general
case where the gauge fields are nonzero and the metric is
arbitrary.

\subsubsection*{Scalars}

The general scalar-field action given above involves a collection
of $N$ real scalar fields, $\Phi^i$, coupled to a background
spacetime metric, $g_{MN}$, and form-fields, $A^a_{M_1..M_{p-1}}$.
Assuming that the scalars couple to a background (1-form) gauge
potential, the background covariant derivative appropriate to this
case is:
\be
    D_M \Phi^i = \partial_M \Phi^i - i A^a_M \, {(t_a)^i}_j \Phi^j
     \,,
\ee
where the matrices ${(t_a)^i}_j$, $i=1 \ldots N$, represent the
gauge group on the scalars.

To compute the one-loop quantum effects of scalar fluctuations we
linearize this action about a particular background configuration,
$\varphi^i$, according to: $\Phi^i = \varphi^i + \phi^i$, where
$\partial_M \varphi = 0$. Expanding the classical action to
quadratic order in $\phi^i$ allows the identification of the
operator ${\Delta^i}_{j}$, which is given by
\beq
    {\Delta^i}_{j} = - {\delta^i}_{j} \, \Box  + {X^i}_{j}
     \,,
\eeq
with ${X^i}_{j}$ given by
\be
    {X^i}_{j} = G^{ik} \Bigl[ V_{kj}(\varphi) + \frac12 \, R \, U_{kj}
    + \frac14 \, F^a_{MN} F_a^{MN} \, W_{kj}(\varphi) \Bigr] \,,
\ee
and
\be
    {(Y_{MN})^i}_j =  -i {F^a}_{MN} {(t_a)^i}_j
    \,.
\ee
In this last expression the subscripts on $U$, $V$ and $W$ denote
differentiation with respect to the background field $\varphi^i$.

Specializing to Ricci-flat geometries with vanishing Maxwell fields,
these simplify to ${X^i}_{j} = G^{ik}\, V_{kj}$ and $Y_{mn} = 0$. We
will also make use of the following result for a Ricci flat space
(we neglect total derivatives)
\be
    D_E R_{ABCD} D^E R^{ABCD} = - I_1 - 4 I_2,
\ee
where $I_1$ and $I_2$ are defined as
\bea
    I_1 &=& R_{ABCD} R^{AB}_{\ \ \ EF} R^{CDEF} \nonumber\\
    I_2 &=& R_{ABCD} R^{A \  C}_{\ E \ F} R^{BEDF}.
\eea
It should be noted that the integrand of the Euler number, $\chi$,
in 6 dimensions is proportional to the combination $(-I_1 + 2 \,
I_2)$.

With these choices, the contributions to the $a_k$ due to a loop
of scalars are given by $N \times N$ matrices, with $a_0$ through
$a_3$ given by
\bea \label{gilkeyscalar}
   a_0 &=& I \nonumber \\
   a_1 &=& - X \nonumber  \\
   a_2 &=& \frac{1}{180} \, \Riem2 + \frac{1}{2} \, X^2 \\
   a_3 &=& -\frac{17}{45360} I_1 + \frac{1}{1620} I_2
   + \frac{1}{360}
   \left( - 60 X^3 -2X \Riem2 + 30 X \Box X \right) \nonumber \\
\eea

For instance, for the particular case $V_{jk} = 0$ we have $X = 0$
and
\bea \label{gilkeyscalarsimp}
    &&(-)^F \, \tr_0[a_0] = N \,, \qquad
    (-)^F \, \tr_0[a_1] = 0 \nonumber \\
    &&(-)^F \, \tr_0[a_2] = \frac{N}{180}  \, \Riem2 \\
    &&(-)^F \, \tr_0[a_3] = -\frac{17 N}{45360}  \, I_1
    + \frac{N}{1620}  \, I_2 \nonumber
\eea

\subsubsection*{Spinors}

For $N$ symplectic-Weyl spinors in 6 dimensions, $\psi^a$ with $a
= 1,...,N$, the action can be taken to be,
\be
     S = - \frac12 \int d^6x \sqrt{-g} \;
     G_{ab}(\varphi) \, \overline\psi^{\;a} \nott{D} \psi^b \,,
\ee
where $\nott{D} = {e_A}^{M} \, \Gamma^A D_M$ with $\Gamma^A$ being
the 6D Dirac matrices and ${e_A}^M$ denoting the inverse of the
sechsbein, ${e_M}^A$, which satisfies ${e_M}^A {e_N}^B \eta_{AB} =
g_{MN}$. Since 6D symplectic-Weyl spinors have 2 complex
components their representation of the 6D Lorentz group has $d =
4$ real dimensions.

The differential operator which governs the one-loop contributions
is in this case $\nott{D} = {e_A}^M \Gamma^A D_M$ and so in order
to use the general results of the previous section we write
(assuming there are no gauge or Lorentz anomalies)
\be
    \log \det \nott{D} = \frac{1}{2} \log\det(-\nott{D}^2)\,,
\ee
which implies
\begin{eqnarray}
\label{eqn: sigma_spin1/2}
    i \Sigma_{1/2} &=& \frac12 \, \Tr \log \nott{D}  =
    \frac{1}{4} \Tr \log \left( -
    {\nott{D}}^2 \right) \nonumber \\
        &=& \frac{1}{4} \Tr \log \left(-\Box - \frac{1}{4} R +
        \frac{i}{2} \Gamma^{AB} F^a_{AB} t_a \right) \,,
\end{eqnarray}
where $\Gamma_{AB} = \frac{1}{2}[\Gamma_A,\Gamma_B]$ and $t_a$
denotes the gauge-group generator acting on the spinor fields.
This allows us to adopt the previous results for the ultraviolet
divergences, provided we divide the result by an overall factor of
2, and use
\be
    X =  -\frac{1}{4} \, R \, I + \frac{i}{2} \,
    \Gamma^{AB} \, F^a_{AB} \, t_a  \,,
\ee
where $I$ is the $\cN \times \cN$ unit matrix, with $\cN = Nd$.
Similarly, we find
\be
    Y_{MN} = -\, \frac{1}{4} \, R_{MNAB} \Gamma^{AB} -i {F^a}_{MN} t_a
    \,,
\ee
and so\footnote{We adopt the convention of using $\Tr[...]$ to
denote a trace which includes the Lorentz and/or spacetime
indices, while reserving $\tr[...]$ for those which run only over
the `flavor' indices which count the fields of a given spin.}
\bea
    \Tr_{1/2}[\y2] &=& - 4 \, \tr_{1/2}(t_a t_b) \, F^a_{MN} F^{bMN}
    -\frac{N}{2} \, \Riem2. \\
\eea

Keeping explicit the sign due to statistics, and dropping terms
which vanish when traced, this leads to the following expressions
for the divergent contributions of $N$ 6D Weyl fermions, where we now
specialize to a Ricci flat background with vanishing gauge fields:
\begin{eqnarray}
    \label{gilkeyspinor}
    &&(-)^F \, \Tr_{1/2}[a_0] = - 2N \,, \qquad
    (-)^F \, \Tr_{1/2}[a_1] = 0,  \nonumber \\
    &&(-)^F \, \Tr_{1/2}[a_2] =  \frac{7N}{720} \, \Riem2  \\
    &&(-)^F \, \Tr_{1/2}[a_3] = - \frac{29N}{45360} \, I_1
    + \frac{N}{648}  \, I_2. \nonumber
\end{eqnarray}

\subsubsection*{Gauge Bosons}

For $N$ gauge bosons, ${\cA}^a_M$, with field strength
$\cF^a_{MN}$ and $a = 1,...,N$, the action is
\be
     S = - \int d^6x \sqrt{-g} \;  \frac14 \, W_2(\varphi)
     \, \cF^a_{MN} \cF_a^{MN}  \,,
\ee
expanded to quadratic order about the background fields: $\cA^a_M
= A_M^a + \delta A^a_M$.

Adding a gauge-averaging term proportional to $\xi^{-1}
W_2(\varphi) \, (D^M \cA_M^a)^2$, it is possible to choose the
gauge parameter $\xi$ such that the one-loop contribution due to
the gauge fields has the standard form, with the differential
operator governing the loop given by
\be
    {\Delta^{aM}}_{bN} = - {\delta^a}_b \, {\delta^M}_N \Box
    + {X^{aM}}_{bN} \,,
\ee
with
\be
    {X^{aM}}_{b N} = -
    {R^M}_{ N} {\delta^a}_{ b} + 2i
    {({\tau}_c)^a}_{b} {F^{c M}}_{N}  \,,
\ee
where $\tau_c$ here denotes a gauge generator in the adjoint
representation.

Since the dimension of the 6-vector representation of the Lorentz
group is $d = 6$, we have $\Tr_V I= 6 N$, $\Tr_V(X) = - N \, R$,
and so
\bea
    \Tr_V(X^2) &=& N\, \Ricci2 + 4 \, C(A) \, \f2 \nonumber\\
    \Tr_V(\y2) &=& - N \, \Riem2 - 6 \, C(A) \, \f2 \,,
\eea
where $C(A)$ is the Dynkin index for $N$ fields in the adjoint
representation: $\tr(\tau_a \tau_b) = C(A) \, \delta_{ab}$. The
subscript `$V$' in these expressions is meant to emphasize that the
trace has been taken over a vector field (as opposed to the physical
spin-1 field, including ghosts).

These expressions suffice to compute $\Tr_V[a_k]$, for the vector
field. Once specialized to Ricci flat geometries with vanishing
background gauge field and constant scalars we find
\begin{eqnarray} \label{gilkeyvector}
 &&(-)^F \, \Tr_V[a_0] = 6 N \,, \qquad
 (-)^F \, \Tr_V[a_1] = 0  \\
 &&(-)^F \, \Tr_V[a_2] = -\frac{N}{20}  \, \Riem2 \nonumber \\
 && (-)^F \, \Tr_V[a_3] = \frac{5N}{1512}  \, I_1
 - \frac{N}{135}  \, I_2. \nonumber
\end{eqnarray}

To this we must add the ghost contribution, which consists of $N$
complex scalar fields having fermionic statistics and transforming
in the adjoint representation of the gauge group. The contributions
to the $a_k$ may be read off from our previously-quoted expressions
for scalar fields in the special case $X = 0$. For such fields we
have
\bea \label{gilkeyghost}
   &&(-)^F \, \Tr_{gh}[a_0] = -2N \,, \qquad
   (-)^F \, \Tr_{gh}[a_1] = 0  \nonumber  \\
   &&(-)^F \, \Tr_{gh}[a_2] = -\, \frac{N}{90} \, \Riem2 \\
   &&(-)^F \, \Tr_{gh}[a_3] = \frac{17N}{22680} \, I_1
   - \frac{N}{810}  \, I_2. \nonumber
\eea

Summing the contributions of eqs.~(\ref{gilkeyvector}) and
(\ref{gilkeyghost}) gives the contribution of $N$ physical 6D
massless gauge bosons:
\begin{eqnarray} \label{gilkeyspin1}
    &&(-)^F \, \Tr_1[a_0] = 4 N \,, \qquad
    (-)^F \, \Tr_1[a_1] = 0 \\
    &&(-)^F \, \Tr_1[a_2] = -\frac{11N}{180}  \, \Riem2
    \nonumber \\
    && (-)^F \, \Tr_1[a_3] = \frac{23N}{5670}  \,I_1 - \frac{7N}{810}
    \,I_2. \nonumber
\end{eqnarray}

\subsubsection*{2-Form Gauge Potentials}

A similar result for a 2-form gauge potential, $B_{MN}$, coupled
only to the background metric may also be obtained, either by
direct calculation \cite{Doug3} or by using results in the
literature \cite{Tseytlin}. The starting point for this is the
action
\be
    S = - \frac{1}{12} \, \int d^6x \, \sqrt{-g} \; W_3(\varphi)
    \, G_{MNL} G^{MNL} \,,
\ee
where $G_{MNL}$ is the field strength, $G = \exd B + \omega_3$ and
$W_3(\varphi)$ is a known function of the background scalar
fields.

In this case it is again possible to choose an appropriate
gauge-averaging term and to keep track of all of the ghosts which
result, along the lines as was done for the gauge potential above.
Once summed over the results obtained in this way for ghosts and
2-form potentials, the result specialized to Ricci-flat geometries
and constant scalar fields is given by
\begin{eqnarray} \label{gilkeyKR}
 &&(-)^F \, \Tr_{2f}[a_0] = 6 N \,, \qquad
 (-)^F \, \Tr_{2f}[a_1] =  0, \\
 &&(-)^F \, \Tr_{2f}[a_2] = \frac{11N}{30}  \, \Riem2 , \nonumber \\
 &&(-)^F \, \Tr_{2f}[a_3] =  - \frac{1193N}{7560}  \, I_1 +
 \frac{17N}{54}  \, I_2 \,, \nonumber
\end{eqnarray}
where $N$ counts the number of 2-form potentials which are
present.

\subsubsection*{Gravitini}

We take spin-3/2 particles to be described by symplectic-Weyl
vector-spinor fields, $\psi_M$, with a lagrangian given by
\beq
    \label{eqn: L_spin3/2}
    \L_{VS} = -\frac{1}{2} \g \, \overline{\psi}_M \Gamma^{MNP}
    D_N \psi_P \,,
\eeq
where $\Gamma^{ABC}=\frac{1}{6}(\Gamma^A \Gamma^B \Gamma^C + \cdots)$
is the totally antisymmetric combination of gamma matrices. In
order to put the spin-3/2 lagrangian into a form for which the
general expressions for the Gilkey coefficients apply, it is
convenient to use the gauge-averaging term
\beq \label{3/2gf}
    \L_{gf} = \frac{1}{2} \sqrt{-g} \, (\overline{\Gamma \cdot \psi})
    \nott{D}(\Gamma \cdot \psi) \,.
\eeq
With this term, and after making the field redefinition $\psi_M
\rightarrow \psi_M - \frac{1}{4} \Gamma_M \Gamma \cdot \psi$, we
find that the lagrangian simplifies to the following,
\beq
 \L_{VS} + \L_{gf} = -\frac{1}{2} \, \sqrt{-g} \,
 \overline{\psi}_M \nott{D}
 \psi^M \,.
\eeq
Thus, the one-loop contribution is
\beq \label{gravitino1loop}
    i \Sigma_{3/2} = \frac{1}{2} \log \det[(\nott{D})^A_{\hspace{5pt}
    B}] = \frac{1}{4}\log\det[(-\nott{D}^2)^A_{\hspace{5pt}
    B}]\,.
\eeq

For a vector-spinor the Lorentz generators are
\beq
    (J_{AB})^C_{\hspace{5pt} D} = -\frac{i}{2} \Gamma_{AB}
    \delta^C_D - i I (\delta^C_A \eta_{BD} - \delta^C_B \eta_{AD}) ,
\eeq
where $I$ is the $\cN \times \cN$ identity matrix, corresponding to
the $\cN = N \, d$ (unwritten) non-vector components of $\psi_M$.
Using the identity $\nott{D}^2 = \Box + \frac{1}{4} [
\Gamma^M,\Gamma^N ][D_M, D_N] $, eq.~\pref{gravitino1loop} can be
put into the required form, with $X$ given by
\beq
    X^{A}_{\hspace{5pt} B} = \left( - \frac{1}{4}R +\frac{i}{2}
    F^{{a}}_{CD} \Gamma^{CD} \, t_{{a}} \right) \,
    \delta^{A}_{B} + \frac{1}{2}
    R^{A}_{\hspace{5pt} BMN} \Gamma^{MN} .
\eeq
For simplicity of notation, we have suppressed writing the various
identity matrices that appear in the above expression.

With the above expression for $X$, and remembering to multiply (as
for the spin-1/2 case) eq.($\ref{eqn: gilkey}$) by an overall factor
of 1/2, we find for Ricci flat backgrounds
\begin{eqnarray} \label{gilkeyVS}
 &&(-)^F \, \Tr_{VS}[a_0] = -12 N \,, \qquad
 (-)^F \, \Tr_{VS}[a_1] =  0, \\
 &&(-)^F \, \Tr_{VS}[a_2] = - \frac{11N}{40}  \, \Riem2 , \nonumber \\
 &&(-)^F \, \Tr_{VS}[a_3] =  - \frac{113N}{7560}  \, I_1 +
 \frac{17N}{540}  \, I_2 \,. \nonumber
\end{eqnarray}

We next consider the contribution from the ghost fields. The
gauge-averaging term eq.~\pref{3/2gf} introduces a complex,
Faddeev-Popov spinor ghost as well as a Nielsen-Kallosh ghost, both
of which must be summed over in order to obtain the result for the
spin-3/2 particle. Since both types of ghosts have the same
lagrangian as the massless spin-1/2 particle considered earlier, we
must add to the vector-spinor result above $(-3)$ times the result
for the spin-1/2 particle. Thus, for the gravitino we find

\bea
    \label{gilkeygravitino}
    &&(-)^F \, \Tr_{3/2}[a_0] = -6 N \nonumber \,, \qquad
    (-)^F \, \Tr_{3/2}[a_1] = 0, \nonumber \\
    &&(-)^F \, \Tr_{3/2}[a_2] = -\frac{73N}{240} \, \Riem2, \\
    &&(-)^F \, \Tr_{3/2}[a_3] = - \frac{197N}{15120}  \, I_1 +
    \frac{29N}{1080}  \, I_2. \nonumber
\eea

\subsubsection*{Gravitons}

Finally, we turn to spin-2 particles. Although it is usually true
that only a single spin-2 particle is massless in any given model,
we include a parameter $N$ which counts the massive spin-2 states.
We do so because there is typically more than one massive spin-2
state in the models of interest, typically arising as part of a
Kaluza-Klein tower or as excited string modes.

The lagrangian for a massless rank-two symmetric field is the
Einstein-Hilbert action,
\beq
    \label{EH}
    \L_{EH} = -\frac{1}{2\kappa^2} \g \, R.
\eeq
Writing the metric as ${g}_{MN} + h_{MN}$, we expand the action
to quadratic order in the fluctuations, $h_{MN}$, about the
background metric, $g_{MN}$. To obtain the required form, we use the
gauge-fixing condition
\beq
    \L_{gf} = -\frac{1}{4 \kappa^2} \g \left( D^M h_{MN} -
    \frac{1}{2} D_N h^A_{\hspace{5pt}A} \right)^2.
\eeq
Finally, writing the resulting lagrangian in terms of the scalar,
$\phi \equiv g_{MN} h^{MN}$, and the traceless symmetric tensor
$\phi_{MN} \equiv h_{MN} - \frac{1}{6}g_{MN} \phi$, we find
\begin{eqnarray} \label{Lscalarrank2}
    &&\frac{8\kappa^2}{\g} ( \L_{EH} + \L_{gf} ) =  \phi^{MN}
    \left[ \Box \phi_{MN} +R \, \phi_{MN} - \left( \phi_{MA}
    R_{N}^{\hspace{7pt} A}+ \phi_{NA}
    R_{M}^{\hspace{7pt} A} \right) \right. \nn \\
    && \left. - (R_{MANB} + R_{MBNA}) \phi^{AB} \right] + \frac{2}{3} \,
    \phi^{MN} R_{MN} \phi - \frac{1}{3} \left( \phi \Box \phi +
    \frac{1}{3}R \, \phi^2 \right).
\end{eqnarray}

The fact that the scalar kinetic term has the wrong sign is
well-known, and can be remedied by the field redefinition $\phi
\rightarrow i \phi$. Notice also that, unlike in 4 dimensions, this
procedure produces a cross-term between $\phi_{MN}$ and $\phi$. This
term will vanish for special choices of background metrics, and in
particular for the Ricci flat case considered here. Specializing
immediately to this situation, we see that the above lagrangian
decouples into a rank-2 symmetric traceless piece and a scalar
piece. However, because $R=0$, the results for the scalar
fluctuations have already been calculated and are given by
eq.~\pref{gilkeyscalarsimp}. From eq.~\pref{Lscalarrank2}, we see
that the rank-2 fluctuations have an $X$ given by
\be \label{Xrank2}
    X^{A \hspace{6pt} B}_{\ M \ N} = R^{A \hspace{6pt} B}_{\ M \ N}
    + R^{B \hspace{6pt} A}_{\ M \ N},
\ee
and so for $N$ rank-2 symmetric traceless fields, we find
\bea
    \label{gilkeysymtr}
    &&(-)^F \, \Tr_{symtr}[a_0] = 20 N \nonumber \,, \qquad
    (-)^F \, \Tr_{symtr}[a_1] = 0, \nonumber \\
    &&(-)^F \, \Tr_{symtr}[a_2] = \frac{17N}{18} \, \Riem2, \\
    &&(-)^F \, \Tr_{symtr}[a_3] = \frac{2309N}{11340}  \, I_1 -
    \frac{166N}{405}  \, I_2. \nonumber
\eea

\begin{table}
\begin{center}
\begin{tabular}{|c|c|c|c|c|}
\hline Field & $(-)^F \, \Tr[a_0]$ & $(-)^F \, \Tr[a_1]$ & $(-)^F
  \, 720 \, \Tr[a_2]$ & $(-)^F \, 45360 \, \Tr[a_3]$ \\
\hline $\phi$   &  $1$ & 0 &    $4 \, R_{ABMN}^{\hspace{30pt} 2}$
&
  $-17 \, I_1 + 28 \, I_2$ \\
$\psi$   & $-2$ & 0 &    $7 \, R_{ABMN}^{\hspace{30pt} 2}$ &
  $-29 \, I_1 + 70 \,I_2$ \\
$A_M$    &  $4$ & 0 &  $-44 \, R_{ABMN}^{\hspace{30pt} 2}$ &
  $184 \, I_1 - 392 \, I_2$ \\
$A_{MN}$ &  $6$ & 0 &  $264 \, R_{ABMN}^{\hspace{30pt} 2}$ &
  $-7158 \, I_1 + 14280 \,  I_2$ \\
$\psi_M$ & $-6$ & 0 & $-219 \, R_{ABMN}^{\hspace{30pt} 2}$ &
  $-591 \, I_1 + 1218 \, I_2$ \\
$g_{MN}$ &  $9$ & 0 &  $756 \, R_{ABMN}^{\hspace{30pt} 2}$ &
  $8919 \, I_1 - 17892 \, I_2$ \\
\hline
\end{tabular}
\end{center}
\caption{Results for massless particles in 6 dimensions, including
ghost contributions and specialized to Ricci-flat background
metrics. Note that the boson fields in this table are real and the
fermion fields are 6D symplectic-Weyl spinors. }
\label{masslessFields}
\end{table}

Next, we consider the ghosts for this field. Since the
gauge-averaging term is $f_M = D^N h_{NM} -\frac{1}{2} D_M
h^A_{\hspace{7pt} A}$, and since the gauge transformations are
$\delta h_{MN} = D_M \xi_N + D_N \xi_M$, we find the following
transformation property
\beq
    \delta f_M = \Box \, \xi_M - R^A_{\hspace{7pt} M} \xi_A \,,
\eeq
leading to a complex spin-one Faddeev-Popov fermionic ghost,
$\omega_M$, with lagrangian
\beq
    \L = - \g \omega_M^{*} ( -\Box \, \delta^M_N + R^M_N) \omega^N.
\eeq
For the Ricci flat case, this has the same form as the vector
lagrangian considered earlier. Thus, the complex ghost will
contribute $(-2)$ times the result of a real vector field,
eq.~\pref{gilkeyvector}. Combining the contributions of the scalar
and the Faddeev-Popov ghost to eq.~\pref{gilkeysymtr}, we obtain the
following results for $N$ physical spin-2 particles,
\bea
    \label{gilkeygraviton}
    &&(-)^F \, \Tr_{2}[a_0] = 9 N \,, \qquad
    (-)^F \, \Tr_{2}[a_1] = 0, \nonumber   \\
    &&(-)^F \, \Tr_{2}[a_2] = \frac{21N}{20} \, \Riem2, \\
    &&(-)^F \, \Tr_{2}[a_3] = \frac{991N}{5040}  \, I_1
    - \frac{71N}{180} \,I_2. \nonumber
\end{eqnarray}
The above results for the Gilkey coefficients for 6D massless
particles are summarized in table~\pref{masslessFields}.

\subsection{6D Massive Particles}

It is relatively easy to compute the result for a massive particle
in 6 dimensions using the above expressions for massless
particles. We may do so because the massive-particle formulae are
obtainable by performing an appropriate sum over the corresponding
massless-particle ones, by virtue of the higher-spin Higgs effect.
We again specialize to vanishing background gauge fields and
Ricci-flat metrics.

For example, for spin-1/2 particles an explicit 6D mass term
necessarily relates a left- and right-handed spinor, and so the
result for a massive spin-1/2 particle is obtained by summing the
results for its constituent left- and right-handed components,
$\psi_\pm$. Since $a_0$ through $a_3$ are the same for both
chiralities, this amounts to multiplying the Gilkey coefficients
by a factor of 2.

Similarly, for massive spin-1 particles it is convenient to use
the covariant Fujikawa-Lee-Sanda \cite{FLS,WbgI&II}
gauge-averaging term, $\xi^{-1} [D^M \cA_M^a + c^a(\varphi)]^2$,
with $c^a(\varphi)$ chosen to remove any scalar-vector cross terms
of the schematic form $\cA_M \partial^M \varphi$. In this case the
gauge fields acquire a mass term which is identical to the mass
term which results for the ghosts and for the would-be Goldstone
bosons fields. We give explicit details of these calculations for
arbitrary dimensions in a companion paper \cite{Doug3}. The upshot
of all of this is that the Gikey-DeWitt coefficients for a massive
spin-1 particle are obtained by summing the results for a massless
spin-1 particle with the result for a real spinless particle with
$X=0$, leading to
\begin{eqnarray} \label{gilkeyspin1m}
    &&(-)^F \, \Tr_{1m}[a_0] = 5 N \,, \qquad
    (-)^F \, \Tr_{1m}[a_1] = 0 \\
    &&(-)^F \, \Tr_{1m}[a_2] = -\frac{10N}{180}  \, \Riem2
    \nonumber \\
    && (-)^F \, \Tr_{1m}[a_3] = \frac{167N}{45360}  \,I_1 - \frac{13N}{1620}
    \,I_2. \nonumber
\end{eqnarray}

A similar argument applies for other higher-spin fields, with the
result being to sum the lower-spin formulae in the following
schematic combinations \cite{Doug3}:
\bea \label{massivefields}
    \phi^m &\ra& \phi \nn \\
    \psi^m  &\ra& \psi_+ + \psi_-\nn \\
    A_M^m  &\ra& A_M + \phi\nn \\
    A_{MN}^m &\ra& A_{MN} + A_M \\
    \psi_M^m &\ra& \psi_{M+} + \psi_{M-} + \psi_+ +\psi_-\nn \\
    g_{MN}^m &\ra& g_{MN} + A_M + \phi . \nn
\eea
In writing these relations we use that the dimension of spacetime
of interest here is even, so that Weyl spinors, $\psi_\pm$, may be
defined.

A check on these relations can also be obtained in an alternative
way. Recall that, for flat space, the little group which preserves
the momentum of a massive particle in $n$ dimensions is $SO(n-1)$,
while that which preserves the standard null momentum of a massless
particle is $SO(n-2)$. For the purposes of calculating ultraviolet
divergences, the counting of states for massive particles in $n$
dimensions is equivalent to the counting for massless particles in
$n+1$ dimensions. Using the more general results for the Gilkey
coefficients of higher spins in $n$ dimensions of
ref.~\cite{Doug3,Tseytlin}, it can be shown that the above counting
of massive states agrees with the result for massless fields in
$n+1$ dimensions. Table~\pref{massiveFields} summarizes the results
which these arguments imply for massive particles in 6 dimensions.

\begin{table}
\begin{center}
\begin{tabular}{|c|c|c|c|c|}
\hline Multiplet & $(-)^F \, \Tr[a_0]$ & $(-)^F \, \Tr[a_1]$ &
  $(-)^F \, 360 \, \Tr[a_2]$ & $(-)^F \, 45360 \, \Tr[a_3]$ \\
\hline
$\phi$   &   $1$ & 0 &    $2 \, R_{ABMN}^{\hspace{30pt} 2}$ &
  $-17 \, I_1 + 28 \, I_2$ \\
$\psi$   &  $-4$ & 0 &   $7 \, R_{ABMN}^{\hspace{30pt} 2}$ &
  $-58 \, I_1 + 140 \, I_2$ \\
$A_M$    &   $5$ & 0 &  $-20 \, R_{ABMN}^{\hspace{30pt} 2}$ &
  $167 \, I_1 - 364 \, I_2$ \\
$A_{MN}$ &  $10$ & 0 &  $110 \, R_{ABMN}^{\hspace{30pt} 2}$ &
  $-6974 \, I_1 + 13888 \, I_2$ \\
$\psi_M$ & $-16$ & 0 & $-212 \, R_{ABMN}^{\hspace{30pt} 2}$ &
  $-1240 \, I_1 + 2576 \, I_2$ \\
$g_{MN}$ &  $14$ & 0 &  $358 \, R_{ABMN}^{\hspace{30pt} 2}$ &
  $9086 \, I_1 - 18256 \, I_2$ \\
\hline
\end{tabular}
\end{center}
\caption{Results for massive particles in 6 dimensions,
specialized to Ricci-flat backgrounds. Note that in this table the
bosonic fields are real and the spinors are symplectic, but not
Weyl.} \label{massiveFields}
\end{table}

\subsection{Supersymmetric Multiplets}

In supergravity theories the ultraviolet sensitivity of the
low-energy theory is often weaker than in non-supersymmetric
models. This weaker sensitivity arises due to the tendency of
bosons and fermions to cancel in loops provided their masses and
couplings are equal. These cancellations may be seen by summing
the above results over the particles appearing in the appropriate
supermultiplets.

The result for the ultraviolet-sensitive part of the one-loop
action obtained by integrating out a supermultiplet is given by
\beq \label{eqn: SGheatkernel}
    \Sigma =  \frac{1}{2} \left( \frac{1}{4 \pi} \right)^{n/2}
    \int d^n x \sqrt{-g} \sum_p (-)^{F(p)}
    \sum_{k=0}^{n/2}  \Gamma ( k - n/2) \, \Tr_p[ m_p^{n-2k}\, a_k ]
    \,,
\eeq
where the sum on $p$ runs over the elements of a supermultiplet.
As is clear from this expression, it is weighted sums of the form
$\sum_p (-)^{F(p)} \, \Tr_p[m_p^{n-2k} \, a_k]$ which control the
UV sensitivity of supersymmetric theories. For example the equal
numbers of bose and fermi states within a supermultiplet
automatically imply
\beq
    \sum_{p\in {\rm sm}} (-)^{F(p)} \Tr_p[a_0] = 0
    \,,
\eeq
which expresses the usual cancellation of the contributions of
bosons and fermions to the cosmological constant if all of the
elements of a supermultiplet share the same mass.

The next few sections record the analogous result for the other
coefficients --- $a_1$ through $a_3$ --- once summed over the
particles within various 6D supermultiplets, under the assumption
that all elements of the supermultiplet have the same 6D mass.
Notice that our discussion of the equivalence between performing
the sum over the UV sensitive part of the 4D contribution of each
KK mode and the UV sensitivity as computed using the full 6D
fields shows that the assumption of equal masses within a 6D
supermultiplet relies only on there being unbroken $(2,0)$
supersymmetry at the high energy scale, $M$, appropriate to the
compactification from higher dimensions down to 6D. Provided that
$M \gg m_{\rm KK} \sim 1/r$, this assumption does {\it not}
require that some supersymmetry remains unbroken below the scale
of the KK masses encountered in the compactification from 6D down
to 4D.

\subsubsection*{Massless 6D Supermultiplets}

We first summarize the particle content of the simplest 6D
supermultiplets, starting first with massless multiplets and then
moving on to massive multiplets. Our discussion follows that of
ref.~\cite{Strathdee}.

Massless multiplets are partially characterized by their
representation properties for the `little group' which preserves
the form of a standard light-like energy-momentum vector. In 5+1
dimensions, the light-like little group contains the rotations,
$SO(4)$, of the 4 spatial dimensions transverse to the direction
of motion of the standard light-like momentum. The representations
of $SO(4) \sim SU(2) \times SU(2)$ corresponding to the simplest
particle types are
\bea
    (\oneb,\oneb): && {\rm scalar} \; (\phi) \nn \\
    (\twob,\oneb),(\oneb,\twob): && {\rm Weyl \; spinors} \; (\psi_\pm) \nn \\
    (\twob,\twob): && {\rm gauge \; potential}  \; (A_M)\nn \\
    (\threeb,\oneb),(\oneb,\threeb): &&
    \hbox{(anti) self-dual 2-form potentials} \; (A^\pm_{MN})\nn \\
    (\threeb,\twob),(\twob,\threeb): && {\rm Weyl \; gravitino}
    \; (\psi^\pm_M) \nn \\
    (\threeb,\threeb): && {\rm graviton} \; (g_{MN}) \nn
\eea
where we denote the particle type by the field with which it is
usually represented. The two representations listed for the Weyl
fermions correspond to the two types of chiralities these fermions
can have. Similarly, the two representations listed for the 2-form
potential correspond to the self- and anti-self-dual pieces, which
are defined to satisfy $G_{MNP} = \pm \, \epsilon_{MNPQRS} \,
G^{QRS}$.

\begin{table}
\begin{center}
\begin{tabular}{|c|rcl|c|}
\hline Multiplet & \multicolumn{3}{|c|}{Representation} & Field Equivalent \\
\hline
 $\fourb$  & $2^2$ &=& (\twob,\oneb;\oneb) +
 (\oneb,\oneb;\twob) & ($\psi$,2$\phi$)  \\
 $\eightb$  & (\oneb,\twob;\oneb) $\times$ $2^2$ &=& (\twob,\twob;\oneb) +
 (\oneb,\twob;\twob) & ($A_M$,2$\psi$) \\
 $\eightb'$  & (\twob,\oneb;\oneb) $\times$ $2^2$ &=&
 (\threeb,\oneb;\oneb) + (\oneb,\oneb;\oneb) +
 (\twob,\oneb;\twob) &
 ($A_{MN}$,2$\psi$,$\phi$) \\
 $\twelveb$ & (\oneb,\threeb;\oneb) $\times$ $2^2$ &=&
 (\twob,\threeb;\oneb) + (\oneb,\threeb;\twob) &
 ($\psi_{M}$,2$A_{MN}$)\\
 $\sixteenb$ & (\twob,\twob;\oneb) $\times$ $2^2$ &=&
 (\threeb,\twob;\oneb) + (\oneb,\twob;\oneb) + (\twob,\twob;\twob) &
 ($\psi_{M}$,2$A_M$,$\psi$) \\
 $\twentyfourb$ & (\twob,\threeb;\oneb) $\times$ $2^2$
 &=& (\threeb,\threeb;\oneb) + (\oneb,\threeb;\oneb) +
 (\twob,\threeb;\twob) &
 ($g_{MN}$,2$\psi_{M}$,$A_{MN}$) \\
\hline
\end{tabular}
\end{center}
\caption{Massless representations of $(2,0)$ supersymmetry in 6
dimensions, labelled by the dimension of the representation of the
corresponding little-group algebra, and by the corresponding 6D
field content. The fermions are taken to be symplectic-Weyl, and
the 2-form potentials are similarly self (or anti-self) dual.}
\label{masslessReps}
\end{table}

Massless supermultiplets are also characterized by the action of
the graded automorphism symmetry which is generated by the
supercharges, which transform as spinors of $SO(4)$. Since the
fundamental spinor representations --- $\twob_+ = (\twob,\oneb)$
and $\twob_- = (\oneb,\twob)$ --- of $SO(4) \sim SU(2) \times
SU(2)$ are pseudo-real, the graded automorphism group in 6
dimensions is $USp(N_+) \times USp(N_-)$, where $N_+$ and $N_-$
(which must both be even) characterize the number of independent,
pseudo-real chiral supersymmetries. The minimal supersymmetry
algebra therefore corresponds to $(N_+,N_-) = (2,0)$, and so $N =
N_+ + N_- =2$ \cite{Strathdee}. The relevant little group
characterizing the massless supermultiplets is then $G = SO(4)
\times USp(2) \sim SU(2) \times SU(2) \times USp(2)$, under which
the active supercharges transform in the representation
$(\twob,\oneb;\twob)$.

\begin{table}
\begin{center}
\begin{tabular}{|c|c|c|c|c|}
 \hline Multiplet & $(-)^F \,
  \Tr[a_0]$ & $(-)^F \, \Tr[a_1]$ & $(-)^F
  \, 48 \, \Tr[a_2]$ & $(-)^F \, 720 \, \Tr[a_3]$ \\
 \hline
  $\fourb$  & 0 & 0 &   $R_{ABMN}^{\hspace{30pt} 2}$ &
  $ (-I_1 + 2 \, I_2)$ \\
 $\eightb$  & 0 & 0 &  $-2 \, R_{ABMN}^{\hspace{30pt} 2}$ &
  $-2 \,(-I_1 + 2 \, I_2)$ \\
 $\eightb'$ & 0 & 0 &  $10 \, R_{ABMN}^{\hspace{30pt} 2}$ &
  $58 \, (-I_1 + 2 \, I_2)$ \\
 $\twelveb$ & 0 & 0 &   $3 \, R_{ABMN}^{\hspace{30pt} 2}$ &
  $123 \, (-I_1 + 2 \, I_2)$ \\
 $\sixteenb$ & 0 & 0 & $-20 \, R_{ABMN}^{\hspace{30pt} 2}$ &
  $4 \, (-I_1 + 2 \, I_2)$ \\
 $\twentyfourb$ & 0 & 0 &  $30 \, R_{ABMN}^{\hspace{30pt} 2}$ &
  $-66 \, (-I_1 + 2 \, I_2)$ \\
 \hline
\end{tabular}
\end{center}
\caption{Results for the statistics-weighted sum over Gilkey
coefficients for massless 6D supermultiplets, specialized to
vanishing gauge fluxes and Ricci-flat backgrounds.}
\label{masslessMultiplets}
\end{table}

The minimal representation of this little algebra has dimension
$2^2 = 4$, and transforms under $G$ like \cite{Strathdee}
\be
    \fourb = (\twob,\oneb;\oneb)+(\oneb,\oneb;\twob).
\ee
This consists of 2 real (1 complex) scalars and a single
symplectic-Weyl fermion, and so consists of 2 bosonic and 2
fermionic states.

Higher-dimensional representations may be obtained from this
minimal one by taking direct products of it with an irreducible
representation of the bosonic part of the little group.
Table~\pref{masslessReps} lists some possible representations
which are obtained in this way, including the hyper-multiplet
($\fourb$), gauge multiplet ($\eightb$), tensor multiplet
($\eightb'$), two types of gravitino multiplet
($\twelveb$, $\sixteenb$) and the graviton multiplet
($\twentyfourb$). To derive the results in this table, we use
the standard results for $SU(2)$: $\twob \times \twob = \oneb +
\threeb$, $\twob \times \threeb = \twob + \fourb$ and $\twob
\times \fourb = \threeb + \fiveb$.

We may now sum the previous expressions for the Gilkey
coefficients over the particle content of these massless 6D
supermultiplets. The results obtained for $\Tr_{\rm sm}[a_k] =
\sum_{p \in {\rm sm}} (-)^{F(p)} \; \Tr_p[a_k]$ if the particles
all share the same mass are summarized in table
\ref{masslessMultiplets}. Notice that the resulting expressions
for $\Tr_{\rm sm}[a_3]$ are proportional to the combination $- I_1
+ 2 I_2$ which gives the Euler number density for compact 6D
manifolds.

\subsubsection*{Massive 6D Supermultiplets}

The massive representations of $(2,0)$ 6D supersymmetry are found
in a similar manner, except in this case the little group for the
time-like energy-momentum vector appropriate to massive fields is
$SO(5)$. The particle types and fields corresponding to the
representations of $SO(5)$ are as follows:
\bea
    \oneb: && \hbox{massive scalar}  \; (\phi^m)  \nn \\
    \fourb: && \hbox{massive spinor} \; (\psi^m) \nn \\
    \fiveb: && \hbox{massive gauge potential} \; (A_M^m) \nn \\
    \tenb: && \hbox{massive 2-form} \; (A^m_{MN}) \nn \\
    \sixteenb: && \hbox{massive gravitino} \; (\psi_M^m) \nn \\
    \fourteenb: && \hbox{massive graviton} \; (g^m_{MN}) \,.\nn
\eea

\begin{table}
\begin{center}
\begin{tabular}{|c|rcl|c|}
 \hline Multiplet & \multicolumn{3}{|c|}{Representation} & Field Equivalent \\
 \hline
 $\sixteenb_m$ & $2^4$ &=& (\fiveb,\oneb) + (\oneb,\threeb) +
  (\fourb,\twob) & ($A_M^m$,$2\psi^m$,3$\phi^m$)  \\
 $\sixtyfourb_m$ & (\fourb,\oneb) $\times$ $2^4$ &=&
  (\sixteenb,\oneb) + (\fourb,\oneb) + (\fourb,\threeb) + &
  ($\psi_M^m$,2$A_{MN}^m$,2$A_M^m$,4$\psi^m$,2$\phi^m$) \\
 &&& (\tenb,\twob) + (\fiveb,\twob) + (\oneb,\twob) & \\
 $\eightyb_m$ & (\fiveb,\oneb) $\times$ $2^4$ &=& (\fourteenb,\oneb)
  + (\tenb,\oneb) + (\oneb,\oneb) + &
  ($g_{MN}^m$,2$\psi_M^m$,$A_{MN}^m$,3$A_M^m$,2$\psi^m$,$\phi^m$) \\
 &&& (\fiveb,\threeb) + (\sixteenb,\twob) + (\fourb,\twob) & \\
 \hline
\end{tabular}
\end{center}
\caption{Massive representations of $(2,0)$ supersymmetry in 6
dimensions, labelled by their dimension. Note that the fermions
are not chiral and the 2-form potentials are not self-dual or
antiself-dual.} \label{massiveReps}
\end{table}

The little algebra for the supersymmetry representations is
therefore $SO(5) \times USp(2)$. The irreducible spinor
representations of $SO(5)$ are not chiral, and are 4-dimensional,
and so the number of supercharges doubles in going from light-like
to time-like representations.\footnote{We assume here vanishing
central charges -- and so no `short' multiplets.} It follows that
the dimensionality of the minimal representation is the square of
what it was in the light-like situation: $2^4 = 16$, with the
following decomposition under $SO(5) \times USp(2)$
\cite{Strathdee}:
\be
    \sixteenb_m = (\fiveb,\oneb) + (\oneb,\threeb) + (\fourb,\twob).
\ee

Again, we find all other representations by taking appropriate
direct products. To do so, we use the following standard results for
$SO(5)$: $\fourb \times \fourb = \tenb + \fiveb + \oneb$ and $\fourb
\times \fiveb = \sixteenb + \fourb$, leading to the massive
supermultiplets given in table \pref{massiveReps}. Since it is
possible to understand massive particles in terms of combinations of
massless ones, the same is true for massive supermultiplets. Using
these results of eq.~\pref{massivefields}, we find the unique
decomposition of the massive multiplets in terms of the massless
ones:
\bea
 \sixteenb_m &\to& 2 \, (\fourb) + \eightb \nn \\
 \sixtyfourb_m &\to& \fourb + 2 \, (\eightb) + 2 \,
  (\eightb') + \twelveb + \sixteenb \label{MassiveMultiplets} \\
 \eightyb_m &\to& 2 \, (\fourb) + \eightb + \eightb'
  + 2 \, (\sixteenb) + \twentyfourb \nn \,.
\eea

Labelling the multiplets as in tables \pref{masslessReps} and
\pref{massiveReps}, it is straightforward to compute the results
for the statistics-weighted sum of the Gilkey coefficients over
the massive supermultiplet particle content. The results obtained
in this way are summarized in Table \ref{massiveMultiplets}. What
is striking about this table is the vanishing of the contributions
to $a_0$ through $a_2$, which ensures the absence within the
effective action of positive powers of the common 6D mass $M$ of a
supermultiplet of degenerate massive particles.
\begin{table}
\begin{center}
\begin{tabular}{|c|c|c|c|c|}
 \hline Multiplet & $(-)^F \, \Tr[a_0]$ & $(-)^F \, \Tr[a_1]$ &
  $(-)^F \, \Tr[a_2]$ & $(-)^F \, \Tr[a_3]$ \\
 \hline
 $\sixteenb_m$ & 0 & 0 & 0 & 0 \\
 $\sixtyfourb_m$ & 0 & 0 & 0 & $\frac{1}{3}(-I_1 + 2 \, I_2)$ \\
 $\eightyb_m$ & 0 & 0 & 0 & 0 \\
 \hline
\end{tabular}
\end{center}
\caption{Results for massive 6D multiplets (Ricci flat).}
\label{massiveMultiplets}
\end{table}

\subsection{Higher-Dimensional Field Content}

Of particular interest are those massive 6D supermultiplets which
are obtained by dimensionally reducing the various 10D
supergravities to six dimensions. The simplest such
compactifications are obtained by reducing the higher-dimensional
theories on a 4-torus. Since the IIA theory can be obtained from
11-dimensional supergravity via dimensional reduction on $S^1$,
the results we obtain for the Type IIA theory dimensionally
reduced on a 4-torus are equivalent to what is obtained from
11-dimensional supergravity dimensionally reduced on a 5-torus.

For the purposes of the present argument all that matters is the
total number of each type of massive 6D fields which are produced
by such a reduction. Our purpose in this section is to show that
the massive field content which is obtained by such a dimensional
reduction is the same as would be obtained by combining a small
number of the massive $\sixteenb_m$, $\sixtyfourb_m$ and $\eightyb_m$
representations of (2,0) 6D supersymmetry.\footnote{This need not
imply that these massive states actually transform in these
representations under (2,0) supersymmetry, such as because of the
possible presence of central charges and short multiplets.} Since
we know that each of these multiplets gives a vanishing
contributions to the first 3 heat-kernel coefficients, the same
must also be true of the contributions of the massive 6D states
which are obtained by dimensional reduction.

\subsubsection*{Type IIA and Type IIB Supergravities}

The field content of Type IIA supergravity in ten dimensions
consists of: a graviton; a 3-form, 2-form, 1-form, and 0-form; two
majorana-Weyl gravitini having opposite 10D chiralities; and two
majorana-Weyl dilatini having opposite 10D chiralities. The IIB
theory is obtainable from the IIA theory by giving all fermions
the same chirality, and by trading the 3-form and the 1-form for a
self-dual 4-form, a 2-form, and a 0-form.

Since each majorana-Weyl spinor in 10D has 16 real components,
reduction on a torus (in the absence of any Scherk-Schwartz
supersymmetry-breaking twists \cite{ScherkSchwarz}) gives $(8,8)$
6D supersymmetry. The massless sector of the resulting 6D theory
can therefore be described in terms of $(2,0)$ supersymmetry
multiplets, and arranges itself into the following collection of
massless $(2,0)$ supermultipets,
\bea
    {\rm Type \, IIA/B}: && \twentyfourb + 4(\sixteenb) +
    2(\twelveb) + 5(\eightb') + 8(\eightb)
    + 10(\fourb) \,,
\eea
with both Type IIA and IIB supergravities giving the same
multiplet content when dimensionally reduced on a 4-torus. In
deriving these results, we use the equivalence (in 6 dimensions)
of a 3-form and a 1-form gauge potential.

Using the results from table~\pref{masslessMultiplets}, we find
the following statistics-weighted sum for the Gilkey coefficients
produced by the massless sector of the 6D theory:
\bea
    &&\Tr[a_0] = \Tr[a_1] = \Tr[a_2] = 0, \nn \\
    &&\Tr[a_3] = \frac{2}{3}(-I_1 + 2I_2).
\eea
The only UV-sensitive quantity is a topological term proportional
to the Euler-number density. If we had not used the duality
relationship to exchange the 3-form potential for a 1-form
potential, even the coefficient of the Euler-number term would
have vanished, because in 6 dimensions
\cite{Tseytlin},\footnote{This is an example of the breakdown of
naive equivalences between different field representation
descriptions of the same particle \cite{Duff}.}
\be
    \Tr_{3f}[a_3] - \Tr_{1f}[a_3] = \frac{2}{3} (-I_1 + 2I_2).
\ee

A similar statement can be made for the massive KK modes produced
by any such $(8,8)$-supersymmetric dimensional reduction on a
4-torus. In this case the massive states have the same field
content as do the following 6D massive $(2,0)$ supersymmetry
representations:
\bea
    {\rm Type \, IIA/B}: && \eightyb_m + 2(\sixtyfourb_m) +
    3(\sixteenb_m) \,,
\eea
The vanishing of $\Tr[a_0]$ through $\Tr[a_2]$ for the massive
multiplets ensures the vanishing of any UV sensitive contributions
from the KK modes obtained when reducing from 10 to 6 dimensions
(provided these are not in short multiplets).

\subsubsection*{Type I and Heterotic Theories}

Type I and heterotic theories consist of a 10D, $N=1$,
supergravity multiplet coupled to a 10D Yang-Mills multiplet. The
relevant gauge group is $SO(32)$ for the type I and $Spin(32)/Z_2$
for the heterotic A theory, and $E_8 \times E_8$ for the heterotic
B theory. The supergravity multiplet contains a graviton, a 2-form
gauge potential, a scalar dilaton, and a majorana-Weyl gravitino
and a majorana-Weyl spin-1/2 fermion, both having the same
chirality. The gauge multiplet contains $N_A$ gauge fields and
$N_A$ spin-1/2 fermions, where $N_A = 496$ is the dimension of the
gauge group.

Dimensionally reducing on a 4-torus to six dimensions without
breaking any supersymmetries in this case ensures unbroken $(4,4)$
6D supersymmetry, with the massless sector arranging itself into
the following $(2,0)$ supermultiplets (where we keep separate the
supergravity and the Yang-Mills fields):
\bea
    {\rm Sugra:} && \twentyfourb + 2(\sixteenb)
    + (\eightb') + 4(\eightb) + 8(\fourb) \nn \\
    {\rm YM:} && N_A (\eightb) + 2 N_A (\fourb). \nn
\eea

Using the results from table~\pref{masslessMultiplets}, we find for
the massless sectors of both the supergravity and the Yang-Mills
theory
\bea
    &&\Tr[a_0] = \Tr[a_1] = \Tr[a_2] = \Tr[a_3] = 0.
\eea
We again find that the dimensionally reduced theory shares the UV
properties of its higher-dimensional counterpart \cite{Tseytlin}.

The contributions of massive KK modes may similarly be analyzed in
this case, with the result that there is a vanishing contribution
to $\Tr[a_0]$, $\Tr[a_1]$ and $\Tr[a_2]$.

\section*{Acknowledgements}

We thank Y. Aghababaie, Z. Chacko, J. Elliot, G. Gabadadze, D.
Ghilencia, A. Tseytlin and F. Quevedo for helpful discussions
about heat-kernel methods and 6D Casimir energies. C.B.'s research
is supported by a grant from NSERC (Canada).

\appendix
\section{Dimensional Regularization and Cutoffs}
\label{DRvsCutoffs}

In this paper all loops are computed using dimensional
regularization since this is by far the most convenient for
practical calculations. At first sight this might appear to limit
our discussion of ultraviolet sensitivity, since this is
traditionally treated in terms of quadratic, quartic and higher
ultraviolet divergences, which are absent in dimensional
regularization. This is a special case of the more general
question of whether (and how) dimensional regularization can be
reconciled with effective field theory techniques, which are
usually phrased in terms of cutoffs, through the integrating out
high-energy modes.

The resolution of this apparent difficulty with dimensional
regularization has been known for quite some time, however, both in
the general context of effective lagrangians
\cite{irrelevant,ETbooks,WeinbergDRUV} and for explicit examples of
UV-sensitivity \cite{UsesAbuses}. The result is that all of the
usual UV-sensitivity issues may also be addressed using dimensional
regularization, and indeed arise there within what is arguably a
more physical context. In this section we briefly recap these
issues, following the discussion of refs.~\cite{UsesAbuses}, in
which the same issues arise within another context.

The issue of UV-sensitivity arises when a physical observable,
${\cal O}$, depends on two very different physical mass scales,
say $m$ and $M$ with $m \ll M$. (The dependence on other
parameters, like couplings {\it etc.} are suppressed in what
follows.) For instance, $M$ and $m$ might be the masses of two
types of particles, which are themselves physically measurable, at
least in principle. It is typically true that the expression for
${\cal O}(m,M)$ simplifies dramatically when $m \ll M$, due to the
ability of expanding the result in powers of the small ratio $m/M$
\be
    {\cal O}(m,M) = m^d \left[ C_{n} \left( \frac{m}{M} \right)^n
    + C_{n-1} \left( \frac{m}{M} \right)^{n-1} + \cdots \right] \,,
\ee
where $d$ is the mass dimension of ${\cal O}$ and the coefficients
$C_k$ might also depend logarithmically on $M/m$. For the vast
majority of observables the starting power, $n$, satisfies $n >
0$, and so the dominant dependence on the large scale, $M$, has
the form $(m/M)^n \log^p(m/M)$ (for some $p \ge 0$) and so
vanishes as $m/M \to 0$ (with the logarithmic dependence arising
through the coefficient, $C_{n}$). UV-sensitivity corresponds to
the case where $n \le 0$, in which case it can be singular to take
the limit $M \to \infty$ in ${\cal O}$ with all other scales
fixed.

Cutoffs arise if the large hierarchy, $M/m$, is exploited to
integrate out all modes having energies $E > \Lambda$ to obtain an
effective theory containing only light particles which is applicable
at scales $E < \Lambda$. Nothing physical depends on the scale
$\Lambda$ in such a construction because it is simply a book-keeping
tool which organizes how calculations are done. This independence
happens in detail through cancellations between the
$\Lambda$-dependence of the effective interactions contained in the
low-energy lagrangian, ${\cal L}$, and the dependence on $\Lambda$
which arises as cutoffs within loop integrals within the low-energy
theory. (See, for instance, refs.~\cite{irrelevant,ETbooks} for
details of this cancellation.)

Within an effective field theory potential confusion can arise
between the dependence of observables, like ${\cal O}$, on
physical heavy masses, like $M$, on the one hand, and the
dependence of the low-energy contribution to ${\cal O}$, on the
cutoff $\Lambda$ on the other. For instance, if a low-energy
calculation of ${\cal O}$ has a low-energy contribution ${\cal
O}_{\rm le} \sim A \, \Lambda^n$, it is tempting to draw the
conclusion that the final result, ${\cal O}$, depends on $M$ like
${\cal O} \sim \hat{A} \, M^n$, for $\hat{A} \sim A$.

This conclusion actually works quite well for logarithmic
UV-sensitivity, because in that case the low energy result, ${\cal
O}_{\rm le} = A \log (\Lambda/m)$, must combine with a high-energy
(effective coupling) contribution, ${\cal O}_{\rm he} = A'
\log(M/\Lambda)$, to give a $\Lambda$-independent result. For
logarithms this dictates that $A = A' = \hat{A}$, so that
\be
    {\cal O} = A \log\left( \frac{\Lambda}{m} \right) + A \log
    \left( \frac{M}{\Lambda} \right) = A \log \left(
    \frac{M}{m} \right) \,,
\ee
and so the coefficient, $A$, of the physical large logarithm,
$\log(M/m)$, may be computed using only the coefficient of the
divergence, $\log(\Lambda/m)$, in the low-energy theory.

The argument can fail, however, for power divergences, for which a
low-energy contribution, ${\cal O}_{\rm le} = A \, \Lambda^2$, can
cancel with a high-energy coefficient, ${\cal O}_{\rm he} = B\,
M^2 - A \, \Lambda^2$, to give the physical answer, ${\cal O} = B
\, M^2$. In this case the coefficient, $B$, of $M^2$ in the full
result need not be related at all to the coefficient, $A$, of the
divergence $\Lambda^2$ within the low-energy theory.

Dimensional regularization and a modified form of modified-minimal
subtraction has many calculational advantages, including the
simplicity of never introducing an unphysical scale like $\Lambda$
as a power during intermediate steps of a calculation. This does
not mean that UV sensitivity cannot arise within a
dimensionally-regulated theory, of course. It merely means that it
arises in a more explicit way - through the matching conditions
which must be applied at a physical threshhold, $\mu = M$, when
one renormalizes the action through this scale and removes the
heavy particles having mass $M$ from the theory
\cite{WeinbergDRUV}. Within this language dangerous powers of $M$
are only obtained when physics at scale $M$ is integrated out in
this way, through explicit contact with the physics of this scale.

In the body of the text we formulate the UV-sensitivity of the
effective 4D cosmological constant in this way, using dimensional
methods to regularize divergent integrations. And it is because of
this choice that we find contributions of order $M^4$ when
integrating out particles of mass $M$, rather than following the
$\Lambda$-dependence in a cutoff regularization.

\section{The Heat Kernel Coefficients}
\label{GilkeyCoefficients}

In this appendix we collect for convenience the explicit expressions
for the coefficients $a_0$ through $a_3$ in their general form for
manifolds without singularities and boundaries. These are known for
general background metrics, $g_{MN}$, and for general $X$ and
$Y_{MN} = [D_M,D_N]$. The first few are given explicitly by
\cite{Gilkey,HKReview}:\footnote{In comparing with this reference
recall that our metric is `mostly plus' and we adopt Weinberg's
curvature conventions \cite{GandC}, which for the Riemann tensor
agree with those of ref.~\cite{HKReview}, but disagree with this
reference by a sign for the Ricci tensor and scalar.}
\begin{eqnarray}
  \label{eqn: gilkey}
  a_0 &=& I \nonumber \\
  a_1 &=& -\frac{1}{6}(RI+6X)  \\
  a_2 &=& \frac{1}{360} \left( 2 \Riem2 - 2 \Ricci2 + 5 R^2 -12\, \Box R \right)
  I \nonumber \\  &&+ \frac{1}{6} R X + \frac{1}{2} X^2 - \frac{1}{6} \Box X +
  \frac{1}{12} \y2 \nonumber\\
  a_3 &=& \frac{1}{7!} \left( - 18 \Box^2 R + 17 D_M R D^M R
  -2 D_L R_{MN} D^L R^{MN}
  -4 D_L R_{MN} D^N R^{ML} \phantom{\frac12} \right. \nonumber\\
  && + 9 D_K R_{MNLP} D^K
    R^{MNLP} +28 R \Box R - 8 R_{MN} \Box R^{MN}
    +24 {R^M}_{N} D^L D^N R_{ML} \nonumber\\
    &&+ 12 R_{MNLP} \Box R^{MNLP}
    - \frac{35}{9} \, R^3 + \frac{14}{3} \, R \,\Ricci2
    - \frac{14}{3} \, R \, \Riem2 \nonumber \\
    && + \frac{208}{9} \, {R^M}_N \, R_{ML} \, R^{NL}
    - \frac{64}{3} \, R^{MN} \, R^{KL} \, R_{MKNL}
    + \frac{16}{3} \, {R^M}_N \, R_{MKLP} \, R^{NKLP} \nonumber\\
    && \left. - \frac{44}{9} \, {R^{AB}}_{MN} \, R_{ABKL}
    \, R^{MNKL} - \frac{80}{9} \, R^{A \ M}_{\ B \ \, N} \,
    R_{AKMP} \, R^{BKNP} \right) I \nonumber \\
    &&+ \frac{1}{360} \left( 8 D_M Y_{NK} \, D^M Y^{NK}
    +2 D^M Y_{NM} \, D_K Y^{NK} + 12 Y^{MN} \Box Y_{MN}
    \phantom{\frac12} \right.  \\
    && - 12 {Y^M}_N \, {Y^N}_K \, {Y^K}_M - 6 R^{MNKL} \, Y_{MN}
    \, Y_{KL} +4 {R^M}_N \, Y_{MK} \, Y^{NK} \nonumber \\
    && - 5 R \, Y^{MN} \, Y_{MN} - 6 \Box^2 X + 60 X \Box X
    +30 D_M X \, D^M X - 60 X^3
    \nonumber \\
    && - 30 X \, Y^{MN} \, Y_{MN} + 10 R \, \Box X + 4 R^{MN}
    \, D_M D_N X +12 D^M R \, D_M X
    -30 X^2 \, R \nonumber \\
    && \left. \phantom{\frac12}
    + 12 X \, \Box R - 5 X \, R^2 + 2 X \, \Ricci2
    -2 X \, \Riem2 \right) \,, \nonumber
\end{eqnarray}
where $I$ is the $N \times N$ identity matrix.

\end{document}